\documentclass[iop]{emulateapj}

\newcommand{\kms}{\mbox{km s$^{-1}$}}
\newcommand{\Msun}{\mbox{$M_{\odot}$}}
\newcommand{\Lsun}{\mbox{$L_{\odot}$}}
\newcommand{\hii}{\mbox{ H\,{\sc ii}}}
\newcommand{\ra}{\mbox{ $\rightarrow$ }}
\newcommand{\x}{\mbox{ $\times$ }}
\newcommand{\HCO}{\mbox{HCO$^{+}$}}
\newcommand{\tHCO}{\mbox{$J=1 \rightarrow0$}}
\newcommand{\NdosH}{\mbox{N$_2$H$^{+}$}}
\newcommand{\tNdosHudt}{\mbox{$JF_1F=123\rightarrow 012$}}
\newcommand{\tNdosHuud}{\mbox{$JF_1F=112\rightarrow 012$}}
\newcommand{\tNdosHucu}{\mbox{$JF_1F=101\rightarrow 012$}}
\newcommand{\HNC}{\mbox{HNC}}
\newcommand{\tHNC}{\mbox{$J=1 \rightarrow0$}}
\newcommand{\CdosH}{\mbox{C$_2$H}}
\newcommand{\tCdosHudcu}{\mbox{$NJF=1\,(3/2)\,2 \rightarrow 0\,(1/2)\,1$}}
\newcommand{\HtreceCO}{\mbox{H$^{13}$CO$^+$}}
\newcommand{\tHtreceCO}{\mbox{$J=1 \rightarrow0$}}
\newcommand{\HNtreceC}{\mbox{HN$^{13}$C}}
\newcommand{\tHNtreceC}{\mbox{$J=1 \rightarrow0$}}
\newcommand{\SiO}{\mbox{SiO}}
\newcommand{\tSiO}{\mbox{$J=2 \rightarrow1$}}
\newcommand{\NHdosD}{\mbox{NH$_2$D}}
\newcommand{\tNHdosD}{\mbox{$J_{K_a,K_c}=1_{1,1} \rightarrow1_{0,1}$}}
\newcommand{\methanol}{\mbox{CH$_3$OH}}
\newcommand{\tCHtresOHf}{\mbox{$J_{K}=2_{-1} \rightarrow1_{-1}$ E}}
\newcommand{\tCHtresOHs}{\mbox{$J_{K}=2_{0} \rightarrow1_{0}$ A}}
\newcommand{\tCHtresOHt}{\mbox{$J_{K}=2_{0} \rightarrow1_{0}$ E}}

\shorttitle{Distinct Chemical Regions in the ``Prestellar'' Infrared Dark
 Cloud (IRDC) G028.23-00.19}
\shortauthors{Sanhueza et al.}

\begin{document}


\title{Distinct Chemical Regions in the ``Prestellar'' Infrared Dark
 Cloud (IRDC) G028.23-00.19}

\author{Patricio Sanhueza\altaffilmark{1}, James M. Jackson\altaffilmark{1},
 Jonathan B. Foster\altaffilmark{2}, Izaskun Jimenez-Serra\altaffilmark{3},
 William J. Dirienzo\altaffilmark{4}, \& Thushara Pillai\altaffilmark{5}}

\altaffiltext{1}{Institute for Astrophysical Research, Boston University, Boston, MA 02215, USA; patricio@bu.edu}
\altaffiltext{2}{Yale Center for Astronomy and Astrophysics, Yale University, New Haven, CT 06520, USA}
\altaffiltext{3}{European Southern Observatory, Karl-Schwarzschild-Str. 2, 85748 Garching, Germany}
\altaffiltext{4}{Department of Astronomy, University of Virginia, P.O. Box 3818, Charlottesville, VA 22903, USA}
\altaffiltext{5}{Caltech, MC 249-17, 1200 East California Blvd, Pasadena, CA 91125, USA}


\begin{abstract}

We have observed the IRDC G028.23-00.19 at 3.3 mm using CARMA. In its center,
 the IRDC hosts one of the most massive ($\sim$1520 \Msun) quiescent, cold
 (12 K) clumps known (MM1). The low temperature, high NH$_2$D
 abundance, narrow molecular linewidths, and absence of embedded IR sources
 (from 3.6 to 70 $\mu$m) indicate that the clump is likely prestellar. Strong
 SiO emission with broad linewidths (6-9 \kms) and high abundances 
 (0.8-4\x10$^{-9}$) is detected in the northern and southern regions of the
 IRDC, unassociated with MM1. We suggest that SiO is released to the gas
 phase from the dust grains through shocks produced by outflows from
 undetected intermediate-mass stars or clusters of low-mass stars deeply
 embedded in the IRDC. A weaker SiO component with narrow
 linewidths ($\sim$2 \kms) and low abundances (4.3\x10$^{-11}$) is detected
 in the center-west region, consistent with either
 a ``subcloud-subcloud'' collision or an unresolved population of a 
 few low-mass stars. We report widespread CH$_3$OH emission throughout the
 whole IRDC and the first detection of extended narrow methanol emission 
($\sim$2 \kms) in a cold, massive prestellar clump (MM1). We suggest that the
 most likely mechanism releasing methanol into the gas phase in such a
 cold region is the exothermicity of grain-surface reactions. HN$^{13}$C
 reveals that the IRDC is actually composed of two distinct substructures
 (``subclouds'') separated in velocity space by $\sim$1.4 \kms. The narrow SiO
 component 
 arises where the subclouds overlap. The spatial distribution of C$_2$H
 resembles that of NH$_2$D, which suggests that C$_2$H also traces cold gas
 in this IRDC.  

\end{abstract}


\newpage

\section{Introduction}

High-mass stars play a key role in the evolution of the energetics and 
chemistry of molecular clouds and galaxies. However, our knowledge of 
their formation is still poor, compared with that of low-mass stars.
 There are several factors that make the study of 
high-mass star formation more challenging than that of their low-mass
 counterparts. High-mass stars are rare and evolve quickly. Massive
 star-forming regions in the earliest stages of evolution are mostly
 located at large distances ($\gtrsim$ 3 kpc) and are heavily extincted
 by dust. The earliest well-understood phase of high-mass star formation
 are the hot molecular cores \citep[HMCs, e.g., ][]{Garay99}. However,
 at this stage, the high-mass stars have already begun burning hydrogen
 and reached the zero-age main sequence (ZAMS), disturbing and changing
 the kinematics and chemistry of the initially starless cocoon. 

Prestellar cores represent the earliest stage of star formation. These 
starless cores are formed by dense, centrally-concentrated parcels of 
gravitationally bound gas \citep[e.g., ][]{Andre09}.
 Prestellar cores will eventually collapse and evolve to form stars.  
The prestellar phase in high-mass star formation has been very 
difficult to find. Initially, large infrared/millimeter surveys had poor
 angular resolution and sensitivity, producing a bias toward bright, evolved
 objects and a misidentification of starless candidates. However,
 with the advent of new space telescopes, such as {\it Spitzer} and
 {\it Herschel}, and submillimeter Galactic plane surveys, it has been
 possible to study in detail the 
best candidates for dense regions which will eventually form massive stars, 
the so-called Infrared Dark Clouds (IRDCs). IRDCs appear as dark 
silhouettes against the Galactic mid-infrared background. Galactic plane
 surveys 
revealed thousands of IRDCs ({\it ISO}, \citealt{Perault96}; {\it MSX},
 \citealt{Egan98}, \citealt{Simon06a}; {\it Spitzer}, \citealt{Peretto09},
 \citealt{Kim10}). The first studies characterizing them suggested that
 they were cold ($<$25 K), massive ($\sim$10$^2$--10$^4$ \Msun), and dense 
($\gtrsim$10$^5$ cm$^{-3}$) molecular clouds with high column densities
 ($\sim$10$^{23}$ cm$^{-2}$) \citep{Carey98,Carey00}. More recent studies
 on IRDC clumps\footnote{Throughout this paper, we use 
the term ``clump'' to refer to a dense object within an IRDC with a size of 
the order $\sim$0.1--1 pc and a mass $\sim$10$^2$--10$^3$ \Msun. We use the 
term ``core'' to describe a compact, dense object within a clump with a size 
$\sim$0.01--0.1 pc and a mass $\sim$1-10$^2$ \Msun.} show that 
they have typical masses of $\sim$120 \Msun, sizes of $\sim$0.5 pc 
\citep{Rathborne06}, dust temperatures that range between 16 and 52 K,  
and luminosities that range from $\sim$10 to 5\x10$^{4}$ \Lsun\ 
 \citep{Rathborne10}. NH$_3$ rotational temperatures are typically lower
 than dust temperatures and range from $\sim$10--20 K
 \citep{Pillai06,Sakai08,Devine11,Ragan11}. 

\begin{figure*}
\begin{center}
\includegraphics[angle=0,scale=0.17]{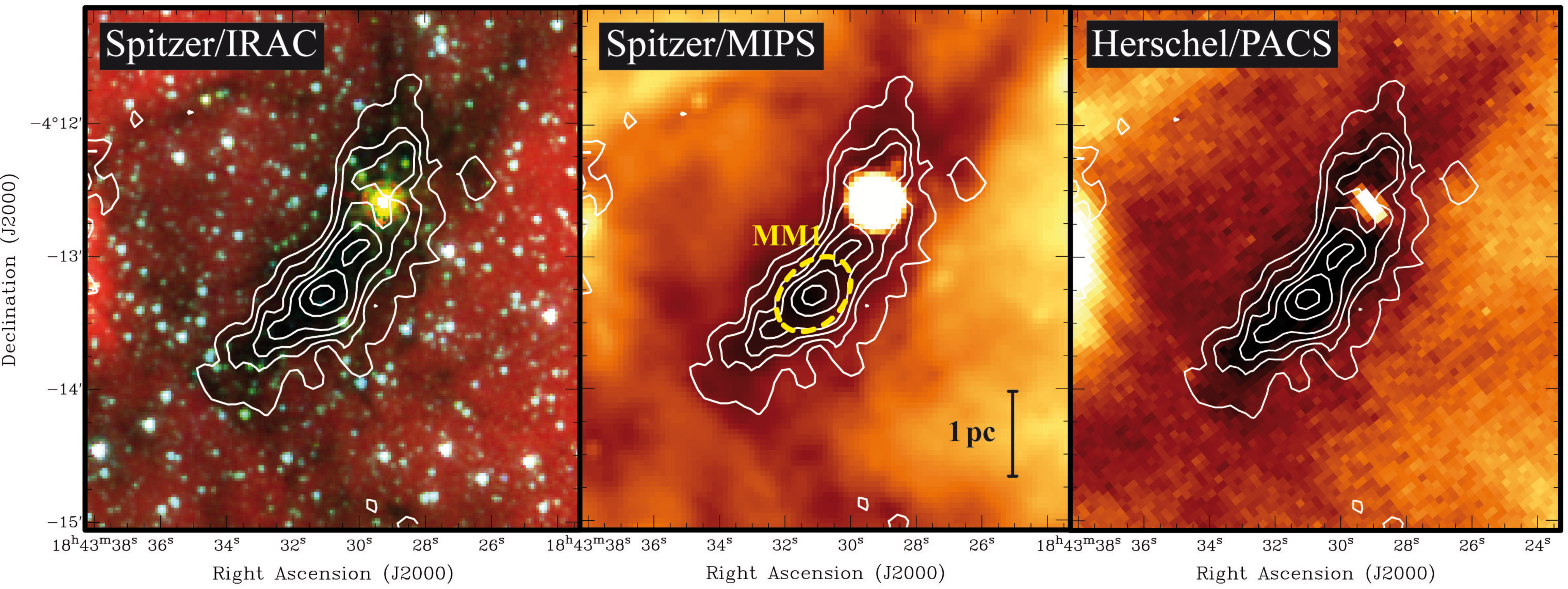}
\end{center}
\caption{ {\it Spitzer} and {\it Herschel} IR images of IRDC G028.23-00.19
 overlaid with 1.2 mm continuum emission from the IRAM 30 m telescope 
(11\arcsec)
 in white contours. Left: {\it Spitzer}/IRAC 3-color (3.6 $\mu$m in blue,
 4.5 $\mu$m in green, and 8.0 $\mu$m in red) image. Center:
 {\it Spitzer}/MIPS 24 $\mu$m image. Right: {\it Herschel}/PACS 70 $\mu$m
 image. Contour levels for the 1.2 mm continuum emission are 20
 ($\sim$3$\sigma$), 35, 50, 65, 85, and 105 mJy beam$^{-1}$. 
Except for the unrelated foreground OH/IR star in the north, there is no
 IR evidence of embedded active star formation.}
\label{IR_image}
\end{figure*}

Evidence of active high-mass star formation in IRDCs is inferred by the
 presence 
of ultracompact (UC) \hii\ regions \citep{Battersby10}, hot cores
 \citep{Rathborne08}, embedded 24 $\mu$m sources \citep{Chambers09},
 molecular outflows \citep{Sanhueza10,Wang11}, or maser emission
 \citep{Wang06,Chambers09}.
IRDCs also host massive, cold IR-dark clumps 
that show similar physical properties to those clumps that host active 
high-mass star formation, except that the temperature and luminosity are
 lower. These characteristics suggest that massive, cold IR-dark clumps
 will form high-mass stars in the future. These objects are the best
 candidates for the most elusive and earliest phase of high-mass star formation,
 the ``prestellar'' or ``starless'' phase. 

Although several studies have focused on IRDCs, the number of studies 
on starless candidates is small \citep[e.g.,][]{Pillai11}. \cite{Zhang09} 
studied in detail an IR quiescent clump, apparently starless, resolving it 
into 5 cores at 1.3 mm with SMA. However, new SMA observations at 0.88 mm 
obtained by \cite{Wang11} show that all cores are associated with molecular
 outflows. This suggests that this clump is in fact forming stars which have
 entered the accretion phase. Thus, it is crucial to evaluate the
 prestellar nature of a core/clump by using interferometric observations
 of molecular lines, and their starless nature should not be asserted just
 by the lack of a detection in the IR.

Little is known about the chemistry of IRDCs. 
A few works have focused on molecular lines surveys of several IRDCs 
\citep{Sakai08,Sakai10,Vasyunina11,Sanhueza12}. For instance, 
 \cite{Sanhueza12} find chemical variations between different evolutionary 
stages using a set of 10 different molecular lines at 3.3 mm. However,
 most of the effort related to chemistry in IRDCs has been focused on 
deuteration. Recent findings suggest that the deuterium
 fraction in high-mass star-forming regions resembles that seen in 
low-mass star-forming regions after the stars turn on, decreasing with time 
as the clumps evolve to warmer temperatures 
 \citep{Chen11,Miettinen11,Fontani11,Sakai12,Pineda13}.  

The internal kinematic structure of IRDCs has not yet been explored in depth.
 In two examples, IRDC G019.30+0.07 \citep{Devine11} and IRDC G035.39-00.33
 \citep{Jimenez10,Henshaw13}, both IRDCs are composed of a few slightly
 different velocity components or ``subclouds''. It is still
 unclear what role these subclouds play in the earliest stages of high-mass
 star formation. 
 
\begin{deluxetable*}{lccccccc}
\tabletypesize{\scriptsize}
\tablecaption{Summary of Observed Molecular Lines and Continuum Emission \label{tbl-obsparam}}
\tablewidth{0pt}
\tablehead{
\colhead{Molecule/} &  \colhead{Molecular}  & \colhead{Frequency}
 & \colhead{$E_u/k$} & \colhead{Beam Size}& \colhead{P.A.} & \multicolumn{2}{c}{\underline {~~~~~~~~~~rms Noise~~~~~~~~~}} \\
\colhead{Continuum} &  \colhead{Transition} & \colhead{(GHz)}&\colhead{(K)}&\colhead{(\arcsec$\times$\arcsec)}&\colhead{(Degrees)}&\colhead{(mJy beam$^{-1}$)}&\colhead{(mK)}\\
}
\startdata
\NHdosD   & \tNHdosD    & 85.926260 & 20.68& 15.3\x8.4 & -25.7 & 33.6 & 43.0\\
\HtreceCO & \tHtreceCO  & 86.754330 & 4.16 & 15.0\x8.5 & -26.9 & 31.1 & 39.8\\
\SiO      & \tSiO       & 86.846998 & 6.25 & 15.0\x8.4 & -27.0 & 31.0 & 39.7\\
\HNtreceC & \tHNtreceC  & 87.090859 & 4.18 & 14.5\x8.6 & -26.1 & 37.2 & 48.4\\
\CdosH    & \tCdosHudcu & 87.316925 & 4.19 & 14.6\x8.6 & -26.3 & 36.4 & 46.6\\
\HCO      & \tHCO       & 89.188526 & 4.28 & 14.6\x8.2 & -27.5 & 33.3 & 42.9\\
\HNC      & \tHNC       & 90.663574 & 4.35 & 14.4\x8.0 & -28.2 & 38.6 & 49.8\\
\NdosH    & \tNdosHuud  & 93.171913 & 4.47 & 14.0\x7.8 & -28.2 & 40.5 & 52.7\\
          & \tNdosHudt  & 93.173772 & 4.47 & 14.0\x7.8 & -28.2 & 40.5 & 52.7\\
          & \tNdosHucu  & 93.176261 & 4.47 & 14.0\x7.8 & -28.2 & 40.5 & 52.7\\
\methanol & \tCHtresOHf & 96.739362 & 12.55& 13.0\x7.6 & -26.0 & 44.5 & 58.8\\
          & \tCHtresOHs & 96.741375 & 6.97 & 13.0\x7.6 & -26.0 & 44.5 & 58.8\\
          & \tCHtresOHt & 96.744550 & 20.10& 13.0\x7.6 & -26.0 & 44.5 & 58.8\\
3.3 mm    & \dots       & 91.877525 & \dots& 14.2\x7.9 & -27.9 &  0.4 & \dots \\
\enddata
\tablecomments{Rest frequencies and upper energy levels were adopted from the 
Splatalogue database (JPL, \citealt{Pickett98}; CDMS,
 \citealt{Muller01,Muller05}), except for N$_2$H$^+$ frequencies which are
 from \cite{Daniel06}.} 
\end{deluxetable*}

 From the studies of \cite{Rathborne10} and \cite{Sanhueza12}, we have
 identified an excellent candidate to study the initial conditions of
 massive star formation. As seen in Figure~\ref{IR_image}, IRDC
 G028.23-00.19 appears to be in a very early stage of evolution because it
 is dark at {\it Spitzer}/IRAC 3.6, 4.5 and 8.0 $\mu$m \citep{Benjamin03}, 
{\it Spitzer}/MIPS 24 $\mu$m \citep{Carey09}, and {\it Herschel}/PACS 70
 $\mu$m \citep{Molinari10}, except for a bright unrelated IR source 
superimposed against the northern part of the cloud. This source is an OH/IR
 star \citep{Bowers89}, and is not a protostar. This foreground low-mass
 star, in a very late stage of evolution, is not associated with the IRDC.
Indeed, its $V_{\rm lsr}$ is $\sim$52 \kms, whereas for IRDC G028.23-00.19 
it is 80 \kms. Located in the center of the IRDC, 
 the clump MM1 is one of the most massive, IR-quiescent clumps known,  
 with a mass of $\sim$640 \Msun, as determined from 1.2 mm continuum emission 
\citep{Rathborne10} at a kinematical distance of 5.1 kpc \citep{Sanhueza12}.
 By fitting the spectral energy distribution (SED), \cite{Rathborne10} 
determined for MM1 a dust temperature of 22 K. Using the Mopra telescope
 (38\arcsec), \cite{Sanhueza12} detected weak molecular 
 line emission of N$_2$H$^+$, HNC and HCO$^+$ $J=1\ra0$ toward this
 clump, consistent with cold gas.  \cite{Battersby10} 
observed this IRDC in 3.6 cm continuum emission with the VLA at 2-3\arcsec\ 
angular resolution and found no \hii\ regions associated with the cloud 
at $\sigma_{\rm rms}=0.01$ mJy beam$^{-1}$, which corresponds to a spectral 
type of B3 or later \citep{Jackson99}. \cite{Chambers09} searched for, 
but failed to detect 22.23 GHz H$_2$O and Class I 24.96 GHz CH$_3$OH masers
 with the GBT at $\sim$32\arcsec\ ($\sigma_{\rm rms}\cong0.05$ K).  
Based on the results obtained from a large variety of observations, we 
suggest that IRDC G028.23-00.19 harbors the prime example of a starless
 or prestellar massive clump (MM1), being in a very early stage of evolution 
and showing no evidence of active star formation.

To probe the chemistry, the internal structure, and the kinematics of IRDCs,
 interferometric observations are critical.
In this paper, we present observations of IRDC G028.23-00.19 in 9 different
 molecular lines obtained with CARMA. We aim to study the chemistry
 and internal structure of a massive star-forming region in a very early
 stage of evolution and test the starless nature of the cloud and,
 in particular, the clump MM1. This is the first study that 
explores the behavior of NH$_2$D, H$^{13}$CO$^+$, SiO, HN$^{13}$C, C$_2$H,
 HCO$^+$, HNC, N$_2$H$^+$, and CH$_3$OH at $\sim$11\arcsec\ angular resolution,
 studying both the chemistry and the structure of multiple subclouds
 traced by distinct, close velocity components ($\lesssim$2 \kms) in a
 massive star-forming region that, apparently, has not yet formed high-mass 
stars. 

\section{Observations}

Observations were carried out with the Combined Array for Research in
 Millimeter-wave Astronomy (CARMA), a 15-element interferometer
 consisting of nine 6.1 m and six 10.4 m antennas, during the CARMA
 Summer School 2011 July 27 in the compact E configuration. The
 projected baselines range from 8 to 66 m. The CARMA correlator has 
eight bands, each with an upper and lower sideband. One band was configured 
to have a bandwidth of 487.5 MHz, resulting in a total bandwidth of $\sim$1 GHz 
by including both sidebands, to observe continuum emission at 3.3 mm for 
calibration (91.878 GHz). The remaining seven bands were used to
 observe 9 different molecular species (NH$_2$D, H$^{13}$CO$^+$, SiO,
 HN$^{13}$C, C$_2$H, HCO$^+$, and HNC in the lower sidebands; and N$_2$H$^+$
 and CH$_3$OH in the upper sidebands). For all seven bands, a bandwidth of
 31 MHz ($\sim$100 \kms) and a spectral resolution of 97 kHz
 ($\sim$0.3 \kms) were used. Using natural weighting, the 1$\sigma$ rms
 noise is 0.4 mJy beam$^{-1}$ for continuum emission, and $\sim$36.2 mJy
 beam$^{-1}$ ($\sim$46.9 mK) per channel for molecular 
 lines. The average of the major and minor axis of the final synthesized 
beam was 10\arcsec.9, which corresponds to a physical size of $\sim$0.27 pc at 
the distance of 5.1 kpc. 
The typical rms noise, synthesized beam, and position angle (P.A.) for the
 continuum emission and each molecular line are given in
 Table~\ref{tbl-obsparam}. 

The system temperature varied from 190 to 250 K during the 7.1 hr track 
(5.4 hr on source).  
The receivers were tuned to a local oscillator (LO) frequency of 91.918 GHz.
 The phase center of the observations is (R.A., Dec.)$_{{\rm J}2000}$ =
 (18$^{\rm h}$43$^{\rm m}$30.$^{\rm s}$78, -04\arcdeg13\arcmin16.\arcsec25).   
At the center frequency of 91.918 GHz, the primary beam or field of view  
of the 6.1 and 10.4 m antennas are 135\arcsec\ and 79\arcsec, respectively.    

The data were reduced and imaged in a standard way using MIRIAD software. 
We used the quasar 1743-038 as a phase calibrator, the quasar 1751+096 as a 
  bandpass calibrator, and Neptune as a flux calibrator.
 The uncertainty in the absolute flux is about 15\%. 

\section{Results}

Figure~\ref{IR_image} shows the {\it Spitzer}/IRAC 3.6, 4.5 and 8.0 $\mu$m
 image (left), 
{\it Spitzer}/MIPS 24 $\mu$m image (center), and {\it Herschel}/PACS 70 $\mu$m
image (right) overlaid with contours of the 1.2 continuum emission (11\arcsec\  
angular resolution).  
As can be seen from the IR images, the IRDC is completely dark, except for the 
foreground OH/IR star superimposed against the northern part of the cloud.
 The 1.2 mm continuum emission,  
from the IRAM 30 m telescope \citep{Rathborne06}, matches the morphology of
 the IR extinction well. The massive clump MM1 is located at 
 the center of the cloud.

\begin{figure*}
\begin{center}
\includegraphics[angle=0,scale=0.22]{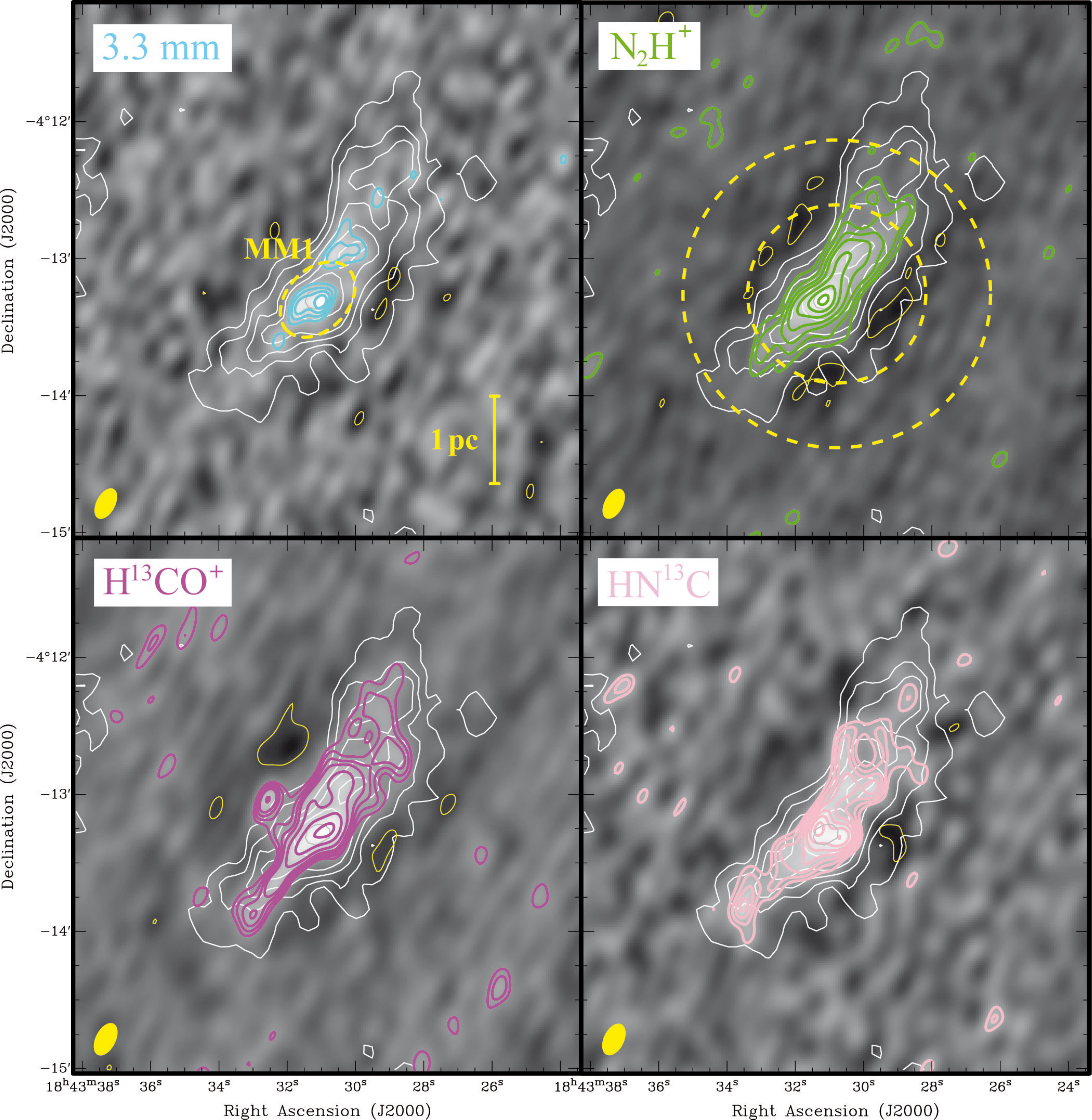}
\end{center}
\caption{\scriptsize \NdosH\ \tNdosHucu, \HtreceCO\ $J=1 \rightarrow0$, and
 \HNtreceC\ $J=1 \rightarrow0$ integrated intensity and 3.3 mm 
 continuum emission maps of IRDC G028.23-00.19 in images and color contours. 
White contours correspond to the 1.2 mm continuum emission from the 
 IRAM 30 m telescope (11\arcsec). Top left: contour levels 
for the 3.3 mm continuum emission are -3, 3, 4, 5, and 6$\sigma$ 
($\sigma=0.4$ mJy beam$^{-1}$). Top right: contour levels for the 
integrated intensity map of \NdosH\ \tNdosHucu, between 76.9 and 82.8 \kms, 
are -4, 3, 5, 7, 9, 12, 15, and 18$\sigma$ ($\sigma=54.0$ mJy beam$^{-1}$
 \kms). Bottom 
left: contour levels for the integrated intensity map of \HtreceCO, between
 76.6 and 82.4 \kms, are -4, 3, 4, 5, 6, 8, 11, and 14$\sigma$ 
($\sigma=43.3$ mJy beam$^{-1}$ \kms). Bottom right: contour levels for
 the integrated
 intensity map of \HNtreceC, between 77.0 and 82.7 \kms, are -4, 3, 4, 5, 6,
 7, 8, and 9$\sigma$ ($\sigma=51.6$ mJy beam$^{-1}$ \kms). Contour levels
 for the 
1.2 mm continuum emission are 20 ($\sim$3$\sigma$), 35, 50, 65, 85, and 105 
mJy beam$^{-1}$. Yellow dashed circles in top right pannel show the 
primary beam of the 6.1 and 10.4 m antennas (135\arcsec\ and 79\arcsec, 
respectively). The synthesized CARMA beam is shown in the bottom left of 
each pannel.}
\label{four_maps1}
\end{figure*}

\begin{figure*}
\begin{center}
\includegraphics[angle=0,scale=0.22]{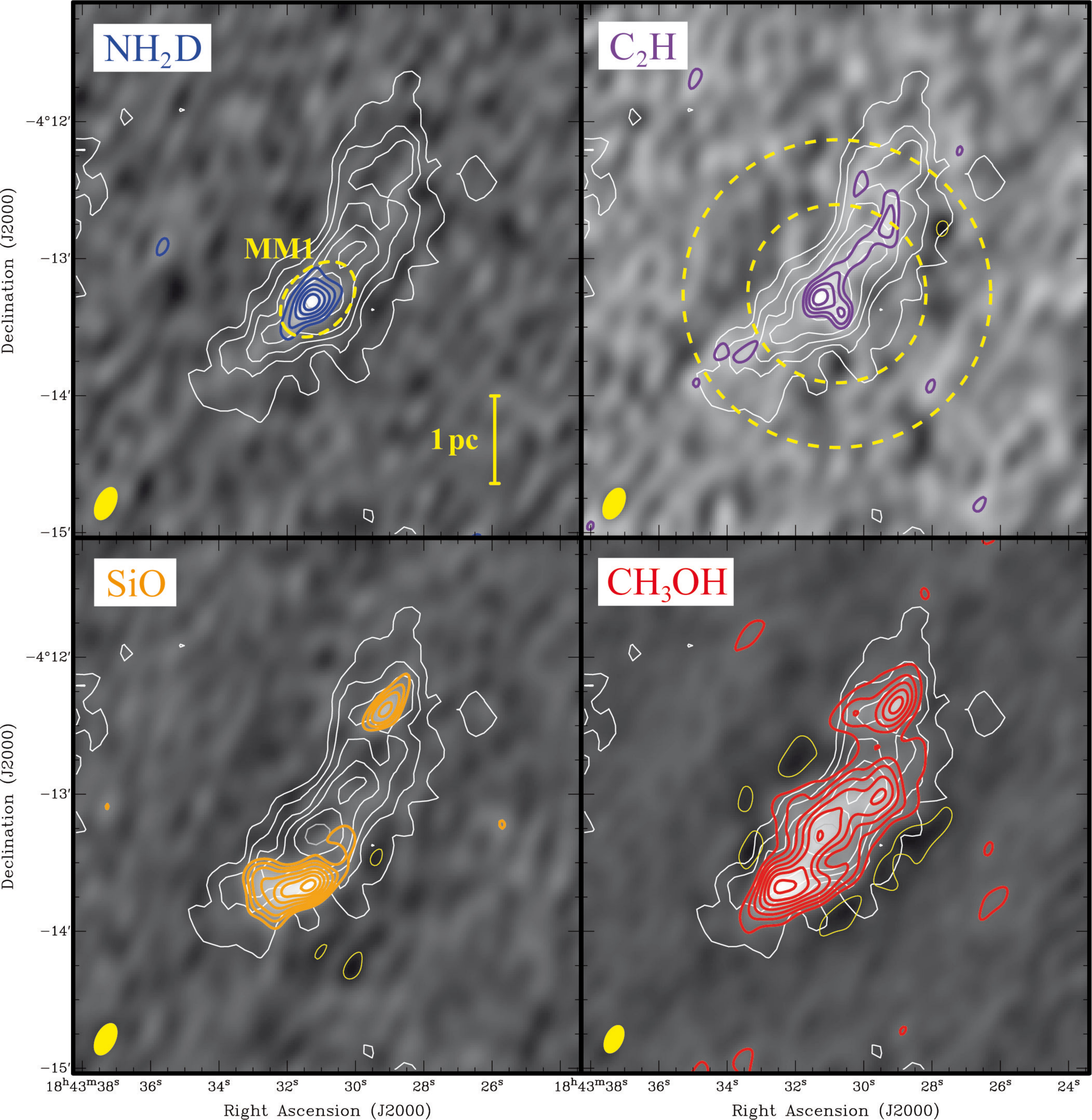}
\end{center}
\caption{\scriptsize \NHdosD\ $J=1 \rightarrow1$, \CdosH\ $J=1 \rightarrow0$,
 SiO $J=2 \rightarrow1$, and \methanol\ $J=2 \rightarrow1$ integrated
 intensity maps of IRDC G028.23-00.19 in images and color contours. 
White contours correspond to the 1.2 mm continuum emission from the 
 IRAM 30 m telescope (11\arcsec). Top left: contour levels for the 
integrated intensity map of \NHdosD, between 78.2 and 80.9 \kms, are -4, 3,
 5, 7, 9, and 11$\sigma$ ($\sigma=32.3$ mJy beam$^{-1}$ \kms). Top right:
 contour levels for the integrated intensity map of \CdosH, between 77.7
 and 82.7 \kms, are -4, 3, 4, 5, and 6$\sigma$ ($\sigma=47.3$ mJy beam$^{-1}$
 \kms). Bottom left: contour levels for the integrated intensity map of
 SiO, between 70.6 and 84.7 \kms, are -4, 4, 6, 8, 11, 14, 17, and 20$\sigma$ 
($\sigma=67.7$ mJy beam$^{-1}$ \kms). Bottom right: contour levels for the
 integrated intensity map of all three \methanol\ transitions, between
 72.9 and 90.8 \kms, are -5, 4, 9, 14, 20, 26, 32, and 40$\sigma$
 ($\sigma=103.6$ mJy beam$^{-1}$ 
 \kms). Contour levels for the 1.2 mm continuum emission are 20
 ($\sim$3$\sigma$), 35, 50, 65, 85, and 105 mJy beam$^{-1}$. Yellow dashed
 circles in top right pannel show the primary beam of the 6.1 and 10.4 m
 antennas (135\arcsec\ and 79\arcsec, respectively). The synthesized CARMA
 beam is shown in the bottom left of each pannel.}
\label{four_maps2}
\end{figure*}

 Figure~\ref{four_maps1} shows the 3.3 mm continuum emission and 
 integrated intensity maps of N$_2$H$^+$ $J=1 \rightarrow0$, H$^{13}$CO$^+$ 
$J=1 \rightarrow0$, and HN$^{13}$C $J=1 \rightarrow0$ in
 images and color contours (ranges of integration are given in the figure 
caption). For N$_2$H$^+$, the transition with the lowest optical depth,
 the isolated hyperfine component \tNdosHucu, was used.   
 White contours are the 1.2 mm continuum emission from IRAM telescope.   
The 3.3 mm continuum emission from CARMA is
 more compact than the 1.2 mm emission. This is probably due to the
 low 3.3 mm flux at 12 K (see Section~\ref{dust_temp}), the low sensitivity 
arising from using only $\sim$1 GHz of continuum bandwidth, and because
 the interferometer may be missing extended flux. In fact, comparing the 
 measured flux at 3.3 mm with the expected 3.3 mm flux from the SED fitting
 done in Section~\ref{dust_temp}, we estimate that the interferometer is
 only recovering about 40\% of the total flux. The emission 
 of N$_2$H$^+$, H$^{13}$CO$^+$, and  HN$^{13}$C resembles
 the 1.2 mm emission from IRAM, although less extended. Using Mopra
 single-pointing observations from 
\cite{Sanhueza12}, we estimate that we recover all the flux for N$_2$H$^+$. 
Except for N$_2$H$^+$, HCO$^+$, and HNC, we have no observations from
 single-dish telescopes 
to evaluate if any other molecular line suffers from missing extended flux. 
Based on the spatial extent of H$^{13}$CO$^+$, HN$^{13}$C, NH$_2$D, C$_2$H, 
SiO, and CH$_3$OH emission, we suggest that zero-spacing problems are not 
important in these line observations and we recover most of the flux with the
 interferometer. On the other hand, using the Mopra single-pointing
 observations from \cite{Sanhueza12}, we estimate that $\sim$30\% of the 
HCO$^+$ and HNC flux is recovered by the interferometer. HCO$^+$ and HNC
 integrated intensity maps are not presented here because their emission
 is substantially resolved out. Therefore, the distribution of these
 molecules is not discussed in the paper. 

\begin{deluxetable*}{lccccc}
\tabletypesize{\scriptsize}
\tablecaption{Gaussian Fit Parameters of Molecular Lines in Selected Positions \label{tbl-gaussian}}
\tablewidth{0pt}
\tablehead{
\colhead{Molecule} &  \colhead{Transition}  & \colhead{$T_{\rm A}$}
 & \colhead{$V_{\rm lsr}$} & \colhead{$\Delta V$}& \colhead{Position} \\
\colhead{} &  \colhead{} & \colhead{(K)}&\colhead{(\kms)}&\colhead{(\kms)}&\colhead{}\\
}
\startdata
\NHdosD   & \tNHdosD    & 0.36 $\pm$ 0.03 & 79.64 $\pm$ 0.05 & 1.34 $\pm$ 0.13 & MM1\\
\HtreceCO & \tHtreceCO  & 0.28 $\pm$ 0.02 & 79.42 $\pm$ 0.09 & 2.41 $\pm$ 0.21 & MM1\\
\HNtreceC & \tHNtreceC  & 0.28 $\pm$ 0.03 & 79.51 $\pm$ 0.09 & 1.89 $\pm$ 0.21 & MM1\\
\CdosH    & \tCdosHudcu & 0.18 $\pm$ 0.03 & 79.28 $\pm$ 0.15 & 1.82 $\pm$ 0.36 & MM1\\
\NdosH    & \tNdosHucu  & 0.66 $\pm$ 0.03 & 79.17 $\pm$ 0.04 & 1.85 $\pm$ 0.10 & MM1\\
\methanol & \tCHtresOHf & 0.89 $\pm$ 0.03 & 79.32 $\pm$ 0.02 & 2.18 $\pm$ 0.09 & MM1\\
          & \tCHtresOHs & 1.12 $\pm$ 0.03 & 79.32 $\pm$ 0.02 & 2.13 $\pm$ 0.07 & MM1\\
          & \tCHtresOHt & 0.13 $\pm$ 0.04 & 79.32 $\pm$ 0.02 & 1.44 $\pm$ 0.47 & MM1\\
\SiO      & \tSiO       & 0.14 $\pm$ 0.02 & 81.79 $\pm$ 0.15 & 1.94 $\pm$ 0.34 & Center-West\\
          &             & 0.26 $\pm$ 0.01 & 76.31 $\pm$ 0.15 & 6.26 $\pm$ 0.36 & South\\
          &             & 0.13 $\pm$ 0.01 & 80.48 $\pm$ 0.33 & 7.74 $\pm$ 0.79 & North\\
\enddata
\end{deluxetable*}

Figure~\ref{four_maps2} shows the integrated intensity maps of NH$_2$D 
$J=1 \rightarrow1$, C$_2$H $J=1 \rightarrow0$, SiO $J=2 \rightarrow1$, and
 CH$_3$OH $J=2 \rightarrow1$ in images and color contours (ranges of
 integration are given in the figure caption). For NH$_2$D, the main component 
$J_{K_a,K_c}=1_{1,1} \rightarrow1_{0,1}$ was used. A velocity range including all
 three transitions was used for CH$_3$OH. The emission of NH$_2$D 
 and C$_2$H is compact and is mostly located in the densest part of the IRDC,
 the MM1 clump. As discussed in Section~\ref{deuterated}, the enhancement 
of these molecules is likely due to cold prestellar gas-phase chemistry. 
Emission from SiO and CH$_3$OH is  
highly unexpected because of the absence of IR signs of star formation (as 
discussed in more depth in Section~\ref{SiO_methanol}). The 
SiO emission is located in the northern and southern regions of the IRDC, and 
it seems to be unassociated with the MM1 clump. An additional weak SiO 
component is located in the center-west part of the cloud. The CH$_3$OH
 emission is widespread over the full IRDC. 

Line center velocities, line widths, and intensity of the lines were 
determined by fitting Gaussian profiles. Molecular lines that show 
hyperfine structures were fitted using a multi-Gaussian function, 
with fixed frequency separation between hyperfine transitions.  
Gaussian fit parameters of molecular lines, in selected positions, are
 shown in Table~\ref{tbl-gaussian}. 

\begin{figure*}
\begin{center}
\includegraphics[angle=0,scale=0.19]{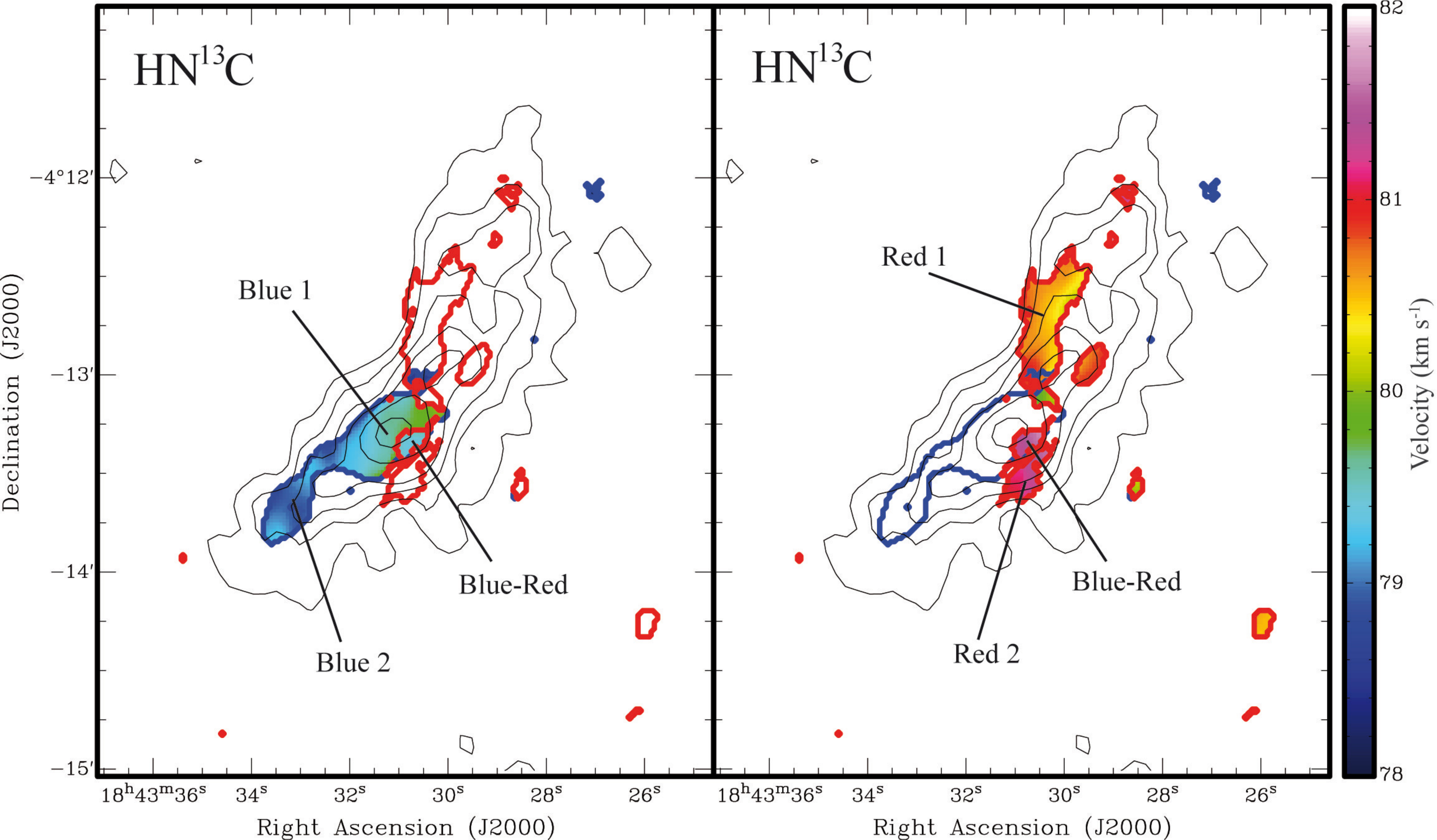}
\includegraphics[angle=0,scale=0.46]{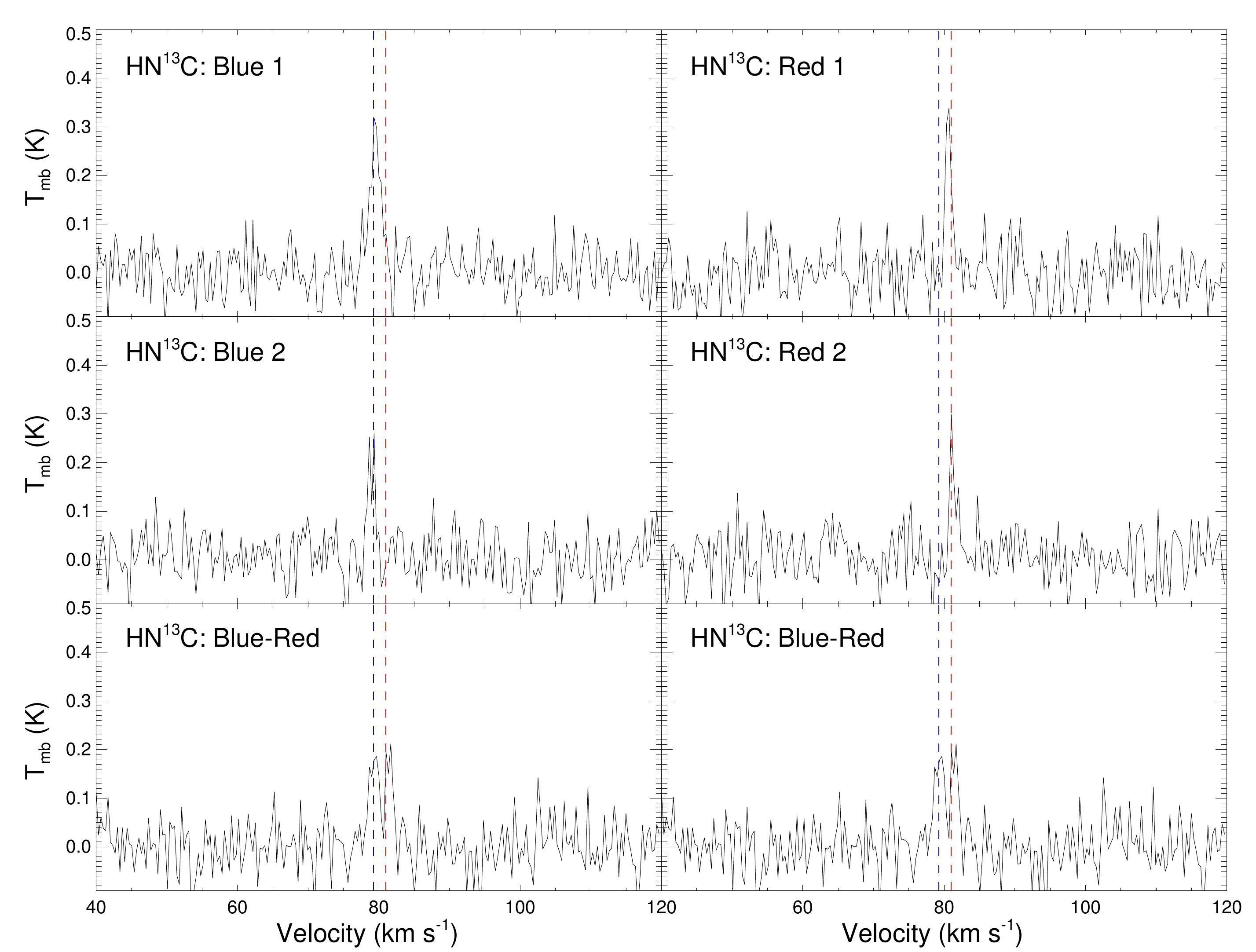}
\end{center}
\caption{\scriptsize Top left: HN$^{13}$C $J=1 \rightarrow0$ central velocity
 map of the blueshifted subcloud (average central velocity of 79.3 \kms). 
Top right: HN$^{13}$C $J=1 \rightarrow0$ center velocity map of the
 redshifted subcloud (average central velocity of 80.7 \kms). Black contours
 correspond to the 1.2 mm continuum emission from the IRAM 30 m telescope.
 Bottom: HN$^{13}$C $J=1 \rightarrow0$ spectra 
 in selected positions. Dashed blue and red lines show the average center 
velocity of the subclouds. Both spectra in the bottom show the same position 
in the cloud where the subclouds overlap.}
\label{hn13c_maps}
\end{figure*}

Multiple velocity components along a line of sight are not unusual in  
star-forming regions. However, it is not easy to separate two velocity
 components with a small offset of $\lesssim$2 \kms\ in massive star-forming
 regions. In general, these kind of regions have line widths  
of 3-4 \kms\ \citep[e.g.,][]{Hoq13}. Two close velocity
 components can 
be separated only when the line emission have narrow profiles and both 
components are not fully blended. This can be done in regions containing
 quiescent gas and using high-density tracers.
 Following the discussion of \cite{Devine11}, we will refer to the structures
 defined by close velocity components as ``subclouds''. 
HN$^{13}$C $J=1 \rightarrow0$ and C$_2$H $J=1 \rightarrow0$ show 
 double-line profiles 
in the center-west part of the IRDC. The double-line profiles are 
actually two close velocity components separated by $\sim$2 \kms.  
Each individual component can be traced continuously from the overlapping 
region of the subclouds, where we see the double-line profiles, to regions 
where only one of the components is seen. Two velocity components are also
 seen in H$^{13}$CO$^+$, N$_2$H$^+$, and CH$_3$OH; however, because these
 molecules have broader line emission, both components are blended.
 Figure~\ref{hn13c_maps} shows the HN$^{13}$C 
 center velocity map of each component obtained by fitting Gaussian profiles
 to the spectra. The map shows two spatially coherent structures. The
 blueshifted subcloud is mostly located in the 
southern-central part of the IRDC and has an average central velocity of 79.3 
\kms. The redshifted subcloud is mostly located in the northern-central
 region of the IRDC and has an average central velocity of 80.7 \kms. The
 velocity difference between subclouds is therefore $\sim$1.4 \kms, similar to
 those found in other IRDCs, such as G019.30+0.07 or G035.39-00.33 
\citep[$<$2 \kms; see][]{Devine11,Henshaw13}. The massive MM1 clump is 
associated with the blueshifted subcloud. NH$_2$D emission is only detected
 in the blueshifted subcloud, while the narrow SiO component (described in the 
next paragraph) is only detected in the redshifted subcloud. All other lines
 are detected in both substructures. 

\begin{figure*}
\begin{center}
\includegraphics[angle=0,scale=0.17]{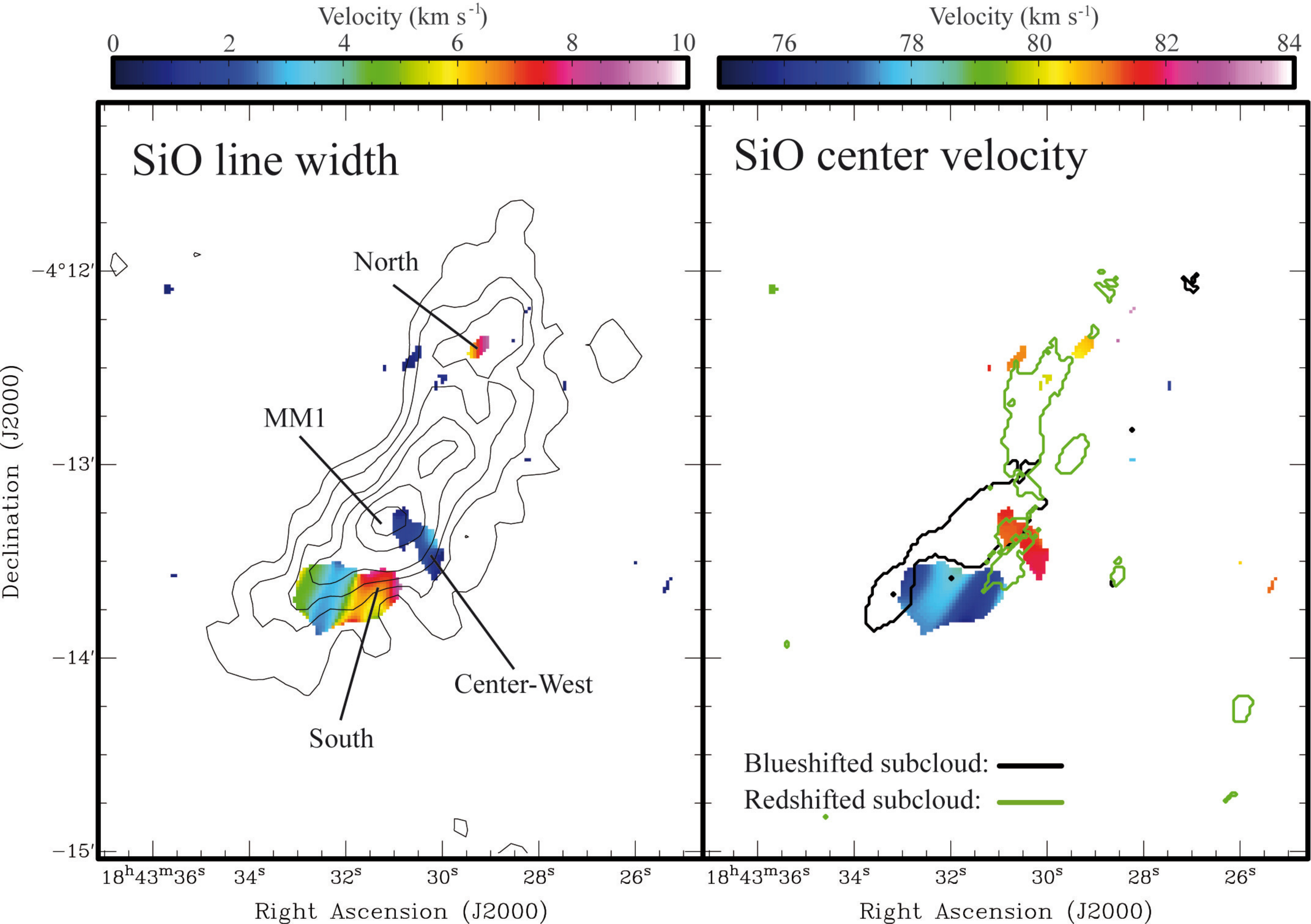}
\includegraphics[angle=0,scale=0.35]{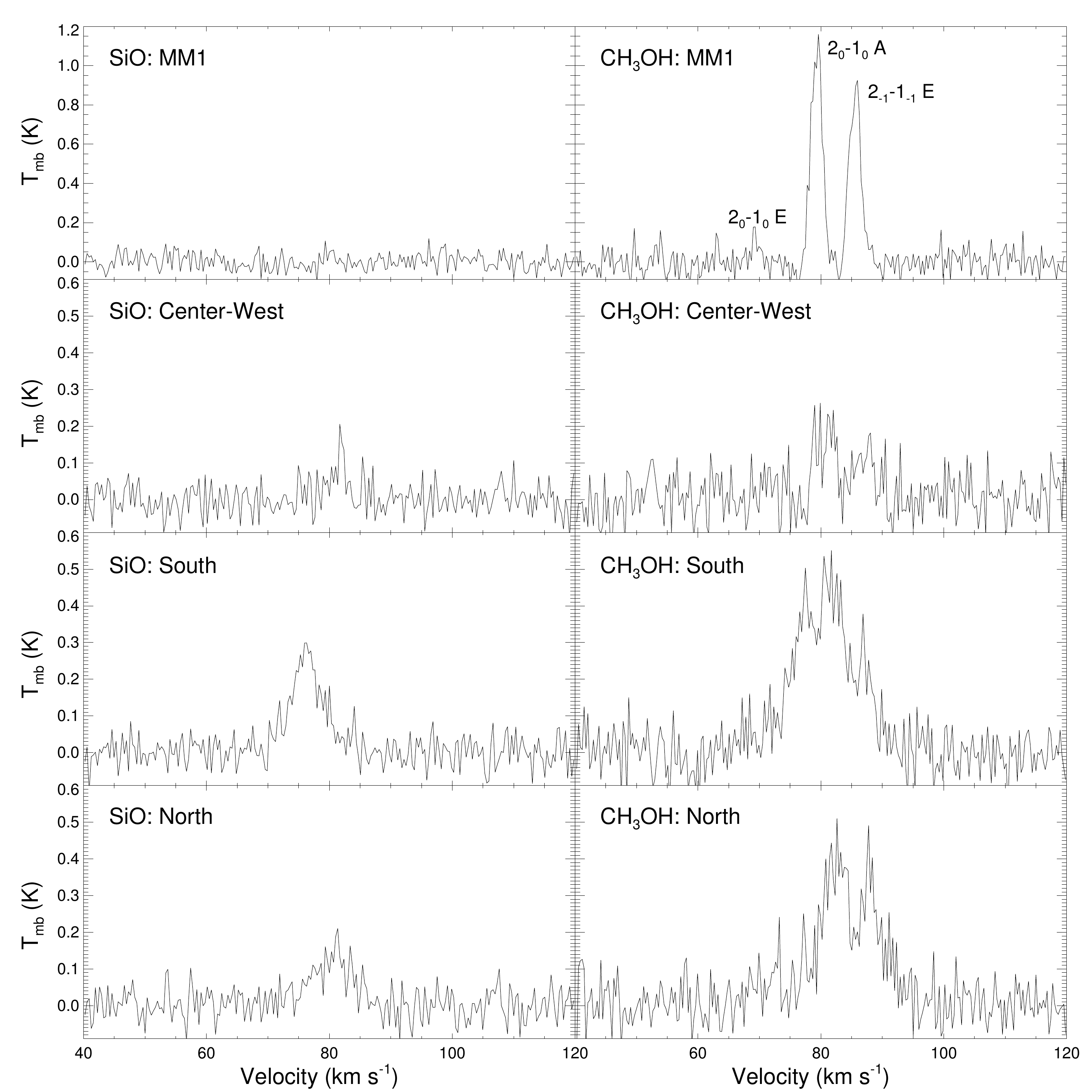}
\end{center}
\caption{\scriptsize Top left: line width velocity map of SiO
 $J=2 \rightarrow1$. Top right: center 
 velocity map of SiO $J=2 \rightarrow1$.  Black contours correspond to
 the 1.2 mm continuum
 emission from the IRAM 30 m telescope. Bottom: spectra of SiO (left) and
 CH$_3$OH (right) in selected positions. In MM1, there is no SiO $J=2
 \rightarrow1$ emission and CH$_3$OH $J=2 \rightarrow1$ shows narrow gaussian
 profiles. In the northern and southern regions of the IRDC, SiO $J=2
 \rightarrow1$ and CH$_3$OH $J=2 \rightarrow1$ lines have broad line
 widths ($\gtrsim$6 \kms). In the center-west part of the IRDC, SiO
 $J=2 \rightarrow1$ lines show narrow ($\sim$2 \kms) and weak profiles. }
\label{sio_methanol}
\end{figure*}

Figure~\ref{sio_methanol} shows line width and center velocity maps of
 SiO across the 
 full IRDC and spectra of SiO and CH$_3$OH in selected positions. Because 
in several positions over the IRDC the CH$_3$OH transitions are blended, 
Gaussian fits can not be performed for all methanol spectra and thus the
 line width and center velocity maps are not useful. The line width map
 of the SiO line reveals the presence of 
two distinct SiO components defined by having different line widths and 
being located in different positions. The first SiO component has narrow
 line widths, $\sim$2 \kms, and peaks at the redshifted velocity component. 
It is also located in the center-west part of the IRDC, at about the same
 position where the subclouds overlap (see right pannel in
 Figure~\ref{sio_methanol}). The second SiO component has broader line 
 widths, $\gtrsim$6 \kms, and is located in the southern and northern
 regions of 
the IRDC. CH$_3$OH emission is also broad in the same positions where
 SiO displays broad line profiles. Figure~\ref{sio_methanol} also displays
 the SiO and CH$_3$OH spectra 
 at the center position of MM1. As expected for quiescent molecular gas
 \citep{Martin92,Requena07}, the SiO emission is not detected at this
 position. The CH$_3$OH 
 \tCHtresOHf\ and \tCHtresOHs\ transitions ($E_u/k$ of 12.6 and 7.0 K,
 respectively) are bright and are not blended. The higher-excitation 
CH$_3$OH \tCHtresOHt\ transition ($E_u/k=20.1$ K) is barely detected
 at 3$\sigma$ level. 

\begin{figure*}
\begin{center}
\includegraphics[angle=0,scale=0.4]{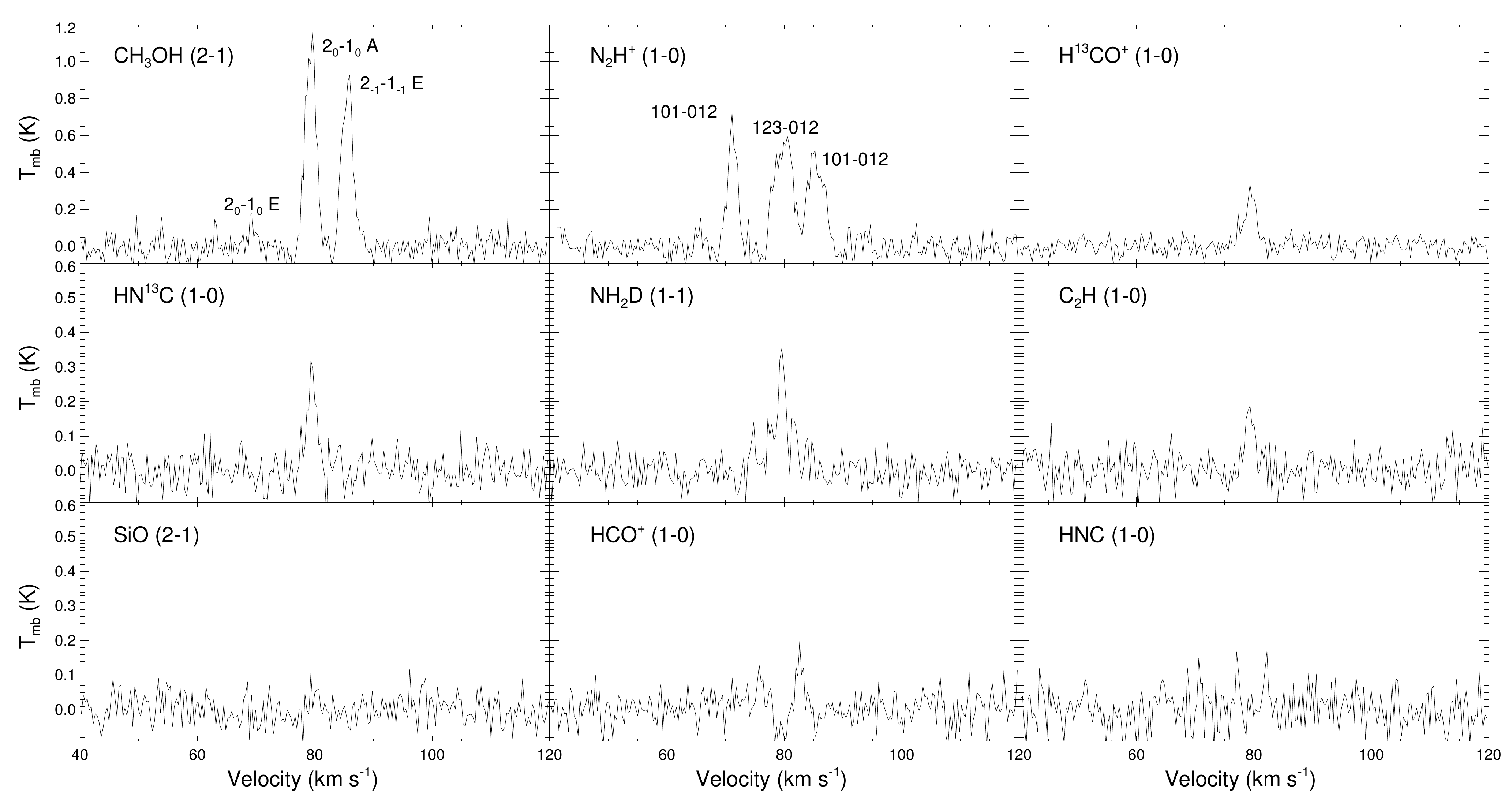}
\end{center}
\caption{Spectra of all nine molecular lines at the peak position of
 NH$_2$D (MM1). Two CH$_3$OH transitions show the brightest emission. The
 N$_2$H$^+$ \tNdosHudt\ and \tNdosHuud\ transitions are saturated and the
 usually weakest \tNdosHucu\ seems to be the brightest line. SiO is not
 detected at this position.}
\label{spec_MM1}
\end{figure*}

Figure~\ref{spec_MM1} shows all lines at the peak position of NH$_2$D (MM1). 
CH$_3$OH is the brightest line at this position. NH$_2$D, C$_2$H, HN$^{13}$C, 
and H$^{13}$CO$^+$ lines show narrow profiles. SiO emission is not detected. 
Because the N$_2$H$^+$ \tNdosHudt\ and \tNdosHuud\ transitions are
 saturated, the usually weakest \tNdosHucu\ is the brightest line, 
suggesting a large optical depth for N$_2$H$^+$ $J=1 \rightarrow0$. HCO$^+$ and 
HNC show only faint emission in MM1. Both HCO$^+$ and HNC show profiles with 
absorption features that prevent their use in calculating physical parameters
 and make their interpretation difficult.  

\section{Discussion}

\subsection{Dust temperature, H$_2$ column density, and mass determination}
\label{dust_temp}

Taking advantage of the new {\it Herschel} Hi-GAL Survey \citep{Molinari10}, 
we derive the dust temperature for MM1 by using the emission detected at
 250, 350, and 500 $\mu$m, and the 1.2 mm emission from IRAM 30 m
 \citep{Rathborne06}. We exclude from the analysis the 70 and 160 $\mu$m
 emission maps. At 70 $\mu$m the IRDC is seen in absorption and at 160 $\mu$m
 the emission from the IRDC is indistinguishable from the
 background/foreground emission. 

After convolving the 250 $\mu$m, 350 $\mu$m, and 1.2 mm maps to the angular 
resolution of the 500 $\mu$m map (35.1\arcsec), we subtract an extended
 component from the {\it Herschel} maps that represents the contribution
 from the  background and foreground diffuse emission. Then, we fit the
 fluxes measured at the peak position of MM1 by using a single temperature
 emission model, assuming optically thin emission, and dust opacities 
($\kappa_{\nu}$) for dust grains with thin ice mantles coagulated at 10$^5$
 cm$^{-3}$ from \cite{Ossenkopf94}. We defer the detailed description of
 the extended emission 
 subtraction algorithm and the fitting procedure to an upcoming paper
 (Guzm\'an et al., in prep.). Using the measured fluxes at 250 $\mu$m (150.0 
MJy ster$^{-1}$), 350 $\mu$m (610.8 MJy ster$^{-1}$), 500 $\mu$m (313.9 MJy
 ster$^{-1}$), and 1.2 mmm (19.5 MJy ster$^{-1}$), we determine a dust 
temperature ($T_{\rm dust}$) of 12 $\pm$ 2 K.  We suggest that the main
 difference 
with the temperature of 22 K determined by \cite{Rathborne10} is that they 
use a low value for the dust emissivity index ($\beta$=1) and use the
 potentially optically thick {\it Spitzer} 85 and 95 $\mu$m emission in
 fitting the spectral energy distribution. The new dust temperature is more
 consistent with the CH$_3$OH temperature we determine by using the
 rotational diagram technique in Section~\ref{rot_temp} (12 K) and the
 NH$_3$ temperature (8 K) recently determined by \cite{Chira13}. 

 To calculate the H$_2$ column density and mass from the 1.2 mm continuum,
 we used the same procedure of \cite{Rathborne06}, assuming a gas-to-dust
 mass ratio of 100 and adopting for the dust opacity ($\kappa_{\rm 1.2\,\,mm}$)
 a value of 1.0 cm$^2$ g$^{-1}$. Computed column densities at a given
 position are shown in Table~\ref{tbl-derived-param}. For the mass, we
 use the same integrated flux measure by \cite{Rathborne10} (1.63 Jy) and
 the distance determined by \cite{Sanhueza12} (5.1 kpc). We obtain a mass
 of 1520 \Msun\ for the clump MM1. 

\begin{deluxetable*}{lccccc}
\tabletypesize{\scriptsize}
\tablecaption{Derived Parameters in Selected Positions\label{tbl-derived-param}}
\tablewidth{0pt}
\tablehead{
\colhead{Molecule} & \colhead{$N_{\rm Molecule}$} & \colhead{$N_{\rm H_2}$}& \colhead{Molecular} & \colhead{Position}\\
\colhead{} & \colhead{(cm$^{-2}$)}&\colhead{(cm$^{-2}$)}&\colhead{Abundance}&\colhead{}\\
}
\startdata
\NHdosD   & 5.2(2.0)\x10$^{13}$ & 3.9(1.4)\x10$^{22}$ & 1.3(0.7)\x10$^{-9}$ & MM1\\
\HCO      & 4.5(0.6)\x10$^{13}$ & 3.9(1.4)\x10$^{22}$ & 1.2(0.5)\x10$^{-9}$ & MM1\\
\HNC      & 5.7(0.9)\x10$^{13}$ & 3.9(1.4)\x10$^{22}$ & 1.5(0.6)\x10$^{-9}$ & MM1\\
\CdosH    & 1.6(0.4)\x10$^{13}$ & 3.9(1.4)\x10$^{22}$ & 4.1(1.8)\x10$^{-10}$& MM1\\
\NdosH    & 1.7(0.2)\x10$^{13}$ & 3.9(1.4)\x10$^{22}$ & 4.3(1.6)\x10$^{-10}$ & MM1\\
\methanol & 1.1(0.1)\x10$^{14}$ & 3.9(1.4)\x10$^{22}$ & 2.7(1.0)\x10$^{-9}$ & MM1\\
\SiO      & 6.7(1.6)\x10$^{11}$ & 1.6(0.5)\x10$^{22}$ & 4.3(1.8)\x10$^{-11}$ & Center-West\\
          & 6.8(0.6)\x10$^{12}$ & 1.8(0.7)\x10$^{21}$ & 3.8(1.4)\x10$^{-9}$ & South\\
          & 8.1(1.2)\x10$^{12}$ & 1.1(0.4)\x10$^{22}$ & 7.5(2.7)\x10$^{-10}$ & North\\
\enddata
\end{deluxetable*}

\subsection{Molecular column densities and abundances}
\label{rot_temp}

Assuming local thermodynamic equilibrium (LTE) conditions, beam-averaged column
 densities of SiO, H$^{13}$CO$^+$, HN$^{13}$C, C$_2$H, and N$_2$H$^+$ were
 calculated using the description of \cite{Sanhueza12}. \cite{Sanhueza12} 
obtain column densities for the same set of lines and list all paramaters 
needed for the calculation. For this work, we have assumed that all lines 
listed above are optically thin. However, there is evidence that the  
 C$_2$H and N$_2$H$^+$ emission lines may be moderately optically thick; 
 their column densities should thus be treated as lower limits.
 HCO$^+$ and HNC column densities are calculated 
 from H$^{13}$CO$^+$ and HN$^{13}$C, adopting a [$^{12}$C/$^{13}$C] isotopic
 abundance ratio of 50 \citep[see discussion of ][]{Milan05}.
 Under LTE conditions, we assume an excitation temperature ($T_{\rm ex}$)
 equal to the dust temperature determined in Section~\ref{dust_temp} (12 K).

For NH$_2$D, the optical depth of the main hyperfine component
 ($\tau_{_{\rm m}}$) was 
 calculated from the ratio between the observed main beam brightness
 temperatures of the main component ($T_{\rm mb_{\rm m}}$) and the brightest
 satellite ($T_{\rm mb_{\rm s}}$), using the  
 following relation \cite[e.g.,][]{Sanhueza12}
\begin{equation}  
\frac{1-e^{-\frac{1}{3}\tau_{\rm m}}}{1-e^{-\tau_{\rm m}}}=\frac{T_{\rm mb_{\rm s}}}{T_{\rm mb_{\rm m}}}~,
\label{N2H+_tau-eqn}
\end{equation}
where we have used that the opacity ratio ($r$) between the main
 ($\tau_{_{\rm m}}$) and  satellite ($\tau_{_{\rm s}}$) lines is
 $r$=$\tau_{_{\rm m}}/\tau_{_{\rm s}}$=3, which depends only on the
 transition moments. We obtained an optical depth of 0.7 for the main
 hyperfine component. 
 
In LTE conditions ($T_{\rm ex}$=$T_{\rm dust}$), the column density is given by
 \cite[e.g.,][]{Garden91,Sanhueza12} 
\begin{equation}
N=\frac{8\pi \nu^3}{c^3R}\frac{Q_{\rm rot}}{g_{\rm u} A_{\rm ul}}\frac{\exp(E_{_
{l}}/kT_{\rm ex})}{[1-\exp(-h\nu/kT_{\rm ex})]}\int \tau \,dv~,  
\label{eqn-col}
\end{equation} 
where $\tau_{\nu}$ is the optical depth of the line, $g_{\rm u}$ is the
 statistical weight of the upper level, $A_{\rm ul}$  
is the Einstein coefficient for spontaneous emission, $E_{_{l}}$  
is the energy of the lower state, $Q_{\rm rot}=3.899+0.751\, T_{\rm ex}^{3/2}$ is
 the partition function for NH$_2$D \citep{Busquet10}, 
$\nu$ is the transition frequency, and $R=1/2$ is
 the relative intensity of the main hyperfine transition with respect
 to the others. $R$ is only relevant for hyperfine transitions because
 it takes into account the satellite lines correcting by their
 relative opacities. It is equal to 1.0 for transitions without
 hyperfine structure. The values of $g_{\rm u}$, $A_{\rm ul}$, and $E_{_{l}}$ are 
given in Table~\ref{tbl-parameters}. The beam-averaged column density
 is obtained by multiplying $N$ by the filling factor, $f$. The
 filling factor can be found from 
\begin{equation}
T_{\rm mb} = f[J(T_{\rm ex}) - J(T_{\rm bg})](1 - e^{-\tau_{\nu}})~,
\label{f}
\end{equation}
 where $J(T)$ is given by 
\begin{equation}
J(T) = \frac{h\nu}{k}\frac{1}{e^{h\nu/kT} - 1}~.
\label{J}
\end{equation}

The derived NH$_2$D column density and filling factor are 6.2$\times$10$^{14}$
cm$^{-2}$ and 0.08, resulting in a beam-averaged column density of
 5.2$\times$10$^{13}$ cm$^{-2}$ in MM1. 

The CH$_3$OH column density ($N$) was obtained by using the rotational
 diagram technique 
 in the most general form \citep[e.g.,][]{Girart02}. Assuming that all
 rotational levels are populated with the same excitation temperature
 (LTE), we have 
\begin{equation}
\log I= \log \frac{fN}{Q_{\rm rot}} - \frac{E_{\rm u}}{T}\log e - \log C_{T} - \log F_{T}~,
\label{I}
\end{equation}
with $Q_{\rm rot}=1.2327\,T^{1.5}$ (the rotational partition function
 for CH$_3$OH) and $I$ equal to 
\begin{equation}
I = \frac{3k\int T_{\rm mb}dv}{8\pi^3g_Kg_IS\mu^2\nu}~,
\end{equation}
where $g_K$ and $g_I$ are the $K-$level and nuclear spin degeneracies, 
$\mu$ is the electric dipole moment of the molecule, and $S$ is the line 
strength. $C_{T}$ and $F_{T}$ are defined as
\begin{equation}
C_{T} = \frac{\tau}{1 - e^{-\tau}}~~~~~{\rm and}~~~~~F_{T} = \frac{J(T_{\rm ex})}{J(T_{\rm ex}) - J(T_{\rm bg})}~.
\end{equation}

The optical depth of the line is given by  
\begin{equation}
\tau = \frac{8\pi^3g_Kg_I\mu^2S\nu}{3h}\frac{N}{Q_{\rm rot}\Delta v}(e^{h\nu/kT} - 1)e^{-E_{\rm u}/kT}~,
\end{equation}
where $E_{\rm u}$ is the energy of the upper state and $\Delta v$ is the
 line width. Values for $g_K$, $g_I$, and $\mu^2S$ are given in
 Table~\ref{tbl-parameters}. A detailed description and derivation of the
 equations used in the rotational diagram technique are presented in
 Silva \& Zhang (2013, in preparation). 

Equation~(\ref{I}) is frequently used without the $C_{T}$ and $F_{T}$ terms, 
assuming that $\tau \ll 1$ and $T_{\rm ex} \gg T_{\rm bg}$. However, in the cold  
clump MM1 these assumptions may not be valid. An additional condition for 
cold gas is that the $E$/$A$ abundance ratio is expected to be 0.69 at low
 temperatures ($\sim$10 K). This is because the ground state of $E$-methanol
 is 7.9 K above the ground state of $A$-methanol, which results in an 
overabundance of $A$-methanol at low temperatures \citep{Friberg88,Wirstrom11}. 
Thus, the integrated intensity of the transition \tCHtresOHs\ was weighted
 by 0.69. Assuming that $f$ (filling factor) equals 1.0, the remaining
 free parameters in Equation~(\ref{I}) are the
 beam-averaged column density and the rotational temperature of CH$_3$OH.
 By fitting the calculated integrated intensities, from Equation~(\ref{I}),
 to the observed integrated intensities of the three different methanol
 transitions, we can find the best solution for the beam-averaged
 column density and the rotational temperature. The fitting procedure was 
 carried out in IDL using the MPFITFUN package \citep{Markwardt09}. At the 
peak position of MM1, the obtained beam-averaged column density and
 rotational temperature are 1.1\x$10^{14}$ cm$^{-2}$ and 12 K. The optical 
depth for the transition \tCHtresOHs\ is 0.1.  

\begin{deluxetable*}{lccccccc}
\tabletypesize{\scriptsize}
\tablecaption{Parameters used for \NHdosD\ and \methanol\ Column Density
 Calculations \label{tbl-parameters}}
\tablewidth{0pt}
\tablehead{
\colhead{Molecule} &  \colhead{Transition}  & \colhead{$g_u$}
 & \colhead{$g_K$} & \colhead{$g_I$}& \colhead{$A_{ul}$} &  \colhead{$E_l/k$} & \colhead{$\mu^2S$}\\
\colhead{} &  \colhead{} & \colhead{}&\colhead{}&\colhead{}&\colhead{($\times10^{-6}$ s$^{-1}$)}&\colhead{(K)}&\colhead{(D$^2$)}\\
}
\startdata
\NHdosD   & \tNHdosD    &   15  & \dots & \dots & 5.8637 & 16.55 & \dots\\
\methanol & \tCHtresOHf & \dots &  2    &  1    & \dots  & \dots & 1.2134\\
          & \tCHtresOHs & \dots &  1    &  2    & \dots  & \dots & 1.6170\\
          & \tCHtresOHt & \dots &  2    &  1    & \dots  & \dots & 1.6166\\
\enddata
\end{deluxetable*}

To estimate the molecular abundances with respect to H$_2$, we take the ratio 
between the column density of a given molecule and the H$_2$ column density 
derived from the 1.2 mm dust continuum emission. We use 
the 1.2 mm continuum for deriving the H$_2$ column density, rather  
the 3.3 mm continuum, because the emission is more extended, more closely
 matches the spatial extent of the molecular line emission, and is detected
 in the regions of interest (for example, where SiO is detected). 
Column densities and abundances for all molecules were calculated for the
 peak positions of NH$_2$D (center of MM1 clump). Additional SiO column
 densities and abundances were calculated for the positions marked in
 Figure~\ref{sio_methanol} (``North,'' ``Center-west,'' and ``South'').

 Uncertainties from the line fitting and dust temperature will propagate to
 the derived column densities and abundances. At the end, uncertainties in 
the molecular abundances range between $\sim$35\% and $\sim$50\%.
 All calculated column densities and abundances with their respective 
uncertainties are summarized in Table~\ref{tbl-derived-param}. 

\subsection{The puzzling SiO and CH$_3$OH emission}
\label{SiO_methanol}

\subsubsection{SiO}

SiO is heavily depleted onto dust grains in quiescent
 regions while it is enhanced in active regions with shocked gas 
\citep{Martin92,Schilke97,Caselli97}. The sputtering and/or grain 
destruction by shock waves injects Si atoms that are rapidly oxidized into
 SiO and/or 
 Si-bearing species that subsequently form SiO molecules in the gas phase.
In the quiescent gas of dark clouds (in low-mass star-forming regions) the
 SiO abundance has an upper limit of $10^{-12}$ 
 \citep[e.g.,][]{Jimenez05,Requena07},
 as expected if the ambient gas has not been significantly affected by
 ejection of material from dust grains. On the other hand, in regions
 with active star formation that have molecular outflows, the SiO abundance 
in the high-velocity gas can be up to 10$^6$ times higher than in the 
quiescent gas \citep[e.g. in L1448,][]{Martin92}. The large variations of 
SiO abundances make this molecule a powerful tracer of shocks. In IRDCs,
 SiO abundances range from 0.1-1\x10$^{-9}$
 \citep{Sanhueza12,Vasyunina11}. Because IRDC G028.23-00.19 appears to lack   
 embedded IR sources that can drive molecular outflows and produce shocks,
 the detection of SiO is highly unexpected. 

The SiO emission in IRDC G028.23-00.19 comprises two different components
 with different kinematics and spatial distribution. This suggests that 
the shocks responsible for releasing  SiO and/or Si-bearing species to the
 gas phase may have 
 two different origins. The SiO component with broad line widths, located 
in the southern and northern regions of the IRDC, could be produced by
 molecular outflows of intermediate-mass stars or clusters of low-mass 
stars, while the SiO component with narrow line widths, located in
 center-west part of the IRDC, could be produced by the collision of
 the subclouds or by the outflows of a few low-mass stars. In the following
 paragraphs, we develop the idea of SiO having different origin in
 different regions of the IRDC.  

In the southern and northern regions, line widths are $\gtrsim$6
 \kms, which are typical values found towards molecular outflows in 
 star-forming regions. Both SiO
 components with broad line widths show no evident wing emission that could
 allow us to identify the blueshifted and redshifted lobes. The molecular
 outflows, if present, could 
  be oriented along the line of sight or the presence of multiple unresolved 
outflows can also make the identification of wings difficult. In the
 southern region, the SiO 
 abundance in the peak position is 3.8$\times$10$^{-9}$. This value is
 typical for regions with molecular outflows \citep[e.g.,][]{Jimenez10,Su07}. 
 In the northern region, the SiO abundance is 7.5$\times$10$^{-10}$. This
 value is at the low end of abundances found in molecular outflows toward 
IRDCs \citep[e.g.,][]{Jimenez10}. Thus, there is evidence supporting the idea 
that the SiO component with broad line widths is produced by molecular 
outflows. However, although the SiO emission in the northern region is
 associated with a weak 1.2 mm peak, the SiO emission in the southern region 
is not associated with any core/clump at mm emission. As determined by fitting 
the SED using {\it Herschel} observations, the bolometric luminosity 
encompassing the northern and southern SiO region are $\sim$300-500 \Lsun\
 (A. Guzm{\'a}n, private communication). Due to the low bolometric
 luminosities, we rule out that the SiO emission in these regions are being
 produced by molecular outflows from deeply embedded high-mass star.
 Based on the values of the line widths, abundances, and luminosities, we
 suggest that the shocks releasing SiO into the gas phase are produced by
 molecular outflows of deeply embedded intermediate-mass stars 
 or clusters of low-mass stars. 

In the center-west region of the IRDC, SiO line widths are significantly
 lower ($\sim$2 \kms) and comparable to other quiescent gas tracers like 
HN$^{13}$C, C$_2$H, and H$^{13}$CO$^+$. This SiO component with narrow line
 widths shows no detectable line wing emission. Not only that, the SiO
 abundance (4.3$\times$10$^{-11}$) is low and only a factor of 10 more
 abundant than the upper limits measured in the quiescent gas of molecular
 dark clouds \citep{Requena07}. These characteristics suggest that molecular
 outflows from intermediate-to-high mass stars may not be causing the
 shocks releasing SiO to the gas phase. The shocks that release this small
 amount of SiO into the gas phase toward the center-west region are
 likely generated by low-velocity shocks
 \citep{Jimenez08,Jimenez10}. As discussed below, these
 shocks may be caused by either the interaction of the subclouds or a
 few low-mass stars.

The SiO component with narrow line widths is remarkably coincident with
 the subcloud-subcloud interface (see right pannel in
 Figure~\ref{sio_methanol}).  As reported by \cite{Jimenez10}, 
 velocities slightly larger than 10 \kms\ are required to sputter SiO from
 the mantles of dust grains in low-velocity C-shocks. However, this velocity
 threshold is above the measured 3D relative velocity between
 subclouds ($\sim$2.4 \kms\ assuming 1D velocity of 1.4 \kms\ in each
 direction). 
 An alternative mechanism for the production of narrow SiO in IRDC 
G028.23-00.19 would be icy mantle vaporization by grain-grain collisions
 similar to what has been proposed for the IRDC G035.39-00.33
 \citep{Caselli97,Henshaw13}. As noted by these authors, ice mantle mixtures
 of H2O:CO have significantly smaller binding energies than pure water ices
 \citep[by a factor of 5;][]{Oberg05}, and therefore require smaller
 vaporization threshold velocities \citep[i.e., ~2.2 \kms, compared to 6.5
 \kms\ for pure water ice; see][]{Tielens94}. The narrow SiO
 emission detected toward the overlapping region between subclouds could
 thus be produced by the vaporization of icy mantles in grain-grain collisions.
 Therefore, the narrow SiO emission would not be associated with star
 formation activity, instead, it would be a signature of a subcloud-subcloud
 collision. We note, however, that the narrow SiO component arises only from
 gas at the velocity corresponding to that one of the redshifted subcloud,
 although this could be a sensitivity effect. 

Alternatively, the SiO component with narrow line widths could be also 
produced by an unresolved population of a few low-mass stars. In this scenario,
 due to the 
distance of the IRDC, beam dilution would prevent us from detecting SiO 
wing emission produced by molecular outflows of the low-mass stars. To 
inspect this scenario, we scaled the SiO emission of a low-mass star-forming
 region located in NGC 1333 (350 pc) to the distance of the 
IRDC G028.23-00.19 (5.1 kpc). Using 1.2 mm continuum emission,
 \cite{Lefloch98a} observed the complex SVS13 and IRAS4A-B, located in NGC
 1333, determining a total mass of 12.6 \Msun\ for the three low-mass cores.
 The angular size of the complex 
 SVS13 and IRAS4A-B at the distance of the IRDC is about the same size
 of the CARMA synthesized beam ($\sim$11\arcsec). \cite{Lefloch98b} observed
 the same region in SiO $J=2\ra1$. Scaling the SiO emission to the
 distance of the IRDC, we find that 1.4 times the emission from a low-mass 
star-forming region like the SVS13 and IRAS4A-B complex could account
 for the SiO emission with narrow 
line widths obtained in one CARMA synthesized beam. Thus, the origin of the SiO 
emission with narrow line widths could be also explained by an unresolved 
population of a few low-mass stars. 

Although higher angular resolution interferometric observations are needed 
to clearly establish the origin of the SiO emission, the component with broad 
SiO line widths and high SiO abundances is consistent with molecular
 outflows from as-of-yet-undetected intermediate-mass stars or 
 clusters of low-mass stars. On the other hand, the component with
 narrow SiO line widths and low SiO abundances is consistent with both 
subcloud-subcloud collision or an unresolved population of a few low-mass
 stars. 

\subsubsection{CH$_3$OH}

Currently, the favored formation path of CH$_3$OH is primarily through 
successive hydrogenation of CO on grain mantles \citep{Watanabe02,Fuchs09}.
 In dark clouds, purely gas-phase reactions produce negligible CH$_3$OH
 abundances \citep[$\lesssim$10$^{-14}$;][]{Garrod06}. Methanol is predominantly 
 formed on grain surfaces and, then in more evolved regions, it is released
 into the gas phase mostly by heating from protostars or sputtering of the
 grain mantles produced by 
 molecular outflows. The study of \cite{Van00} found three types of CH$_3$OH 
abundances in massive star-forming regions: $\sim$10$^{-9}$ for 
the coldest sources, between 10$^{-9}$ and 10$^{-7}$ for warmer sources, 
 and $\sim$10$^{-7}$ for hot cores. Methanol has also been observed in 
IRDCs that show signs of star formation 
 \citep[e.g.,][]{Sakai10,Sakai08,Gomez11}. Because IRDC G028.23-00.19
 has no detection of embedded IR sources that can heat the environment
 or drive molecular 
 outflows, the detection of CH$_3$OH is highly unexpected. This is the 
first detection of extended methanol emission in a massive starless clump,
 although it has been previously detected in low-mass starless cores
 \citep[e.g. in TMC-1 and L134N;][]{Friberg88,Dickens00}.

As in the case of SiO, CH$_3$OH emission also presents two different
 components with different kinematics and spatial distributions. This 
suggests that there may be two different mechanisms releasing methanol to the 
gas phase. The CH$_3$OH emission with broad line widths, co-located 
with the SiO emission with broad line widths, could be produced by   
molecular outflows. The CH$_3$OH emission with narrow
 line widths, located in the center region of the IRDC (including MM1 and 
the SiO region with narrow line widths), could be produced by a non-thermal 
mechanism (e.g., cosmic ray, UV-photons), other than shocks. We discard the
 notion that thermal evaporation 
 of methanol can occur anywhere in IRDC G028.23-00.19 based on the 
non-detection of embedded IR sources that could heat the environment.
 CH$_3$OH evaporates at higher temperatures
 \citep[$\sim$80-100 K,][]{Brown07,Green09} than those measured in this IRDC.
 Although a protostar could exist deeply embedded in the cloud 
 and be undetected in the current IR observations, the methanol emission
 is widespread and the effective heating area of such a protostar would be too 
small to account for the spatial extension of the CH$_3$OH emission.

In the southern and northern regions of the IRDC, where both CH$_3$OH and 
SiO have broad lines, CH$_3$OH line widths are $\gtrsim$4 \kms. Because the
 methanol transitions are blended, we cannot fit the lines and use the 
rotational diagram technique to determine column densities. 
 We conclude that in these regions 
 CH$_3$OH column densities and abundances are higher than in MM1 
(i.e., greater than 1.1\x10$^{14}$ cm$^{-2}$ and 2.7\x10$^{-9}$) based on
 the comparison of the integrated intensities and apparently larger 
 optical depths. Large line widths and abundances may indicate that in the 
southern and northern regions of the IRDC, CH$_3$OH has been released to
 the gas phase by outflow activity as, for example, is occurring in
 IRDC G11.11-0.12 \citep{Gomez11}.
 
On the other hand, the methanol emission that emanates from the center 
of the IRDC can not be explained by molecular outflows. The line widths are 
$\sim$2 \kms, typical of regions without outflow activity. The methanol
 abundance of 2.7\x10$^{-9}$ is about 1-2 orders of magnitude 
lower than that observed in massive star-forming regions with molecular 
outflows and 1 order of magnitude lower than that produced by low velocity 
shocks \citep{Jimenez05}. 
From all the non-thermal mechanisms (e.g., cosmic ray, UV-photons) that
 could explain the observed methanol abundances in the central region of 
IRDC G028.23-00.19, the 
 exothermicity of surface addition reactions is likely the process that
 better reproduces the observed abundances. In this mechanism,
introduced by \cite{Garrod06} and further investigated by \cite{Garrod07},
 the chemical energy released from the grain-surface addition
 reactions is able to break the methanol-surface bound, yielding methanol
abundances of 3-4$\x10^{-9}$, similar to those measured toward the massive, 
quiescent clump MM1. The high extinction found toward IRDC G028.23-00.19,
 and the lack of internal heating sources, makes the non-thermal UV
 photo-desorption of CH$_3$OH unlikely. Cosmic ray-induced desorption, a
 physical mechanism frequently used in dark cloud models
 \citep[e.g.,][]{Hasegawa93}, cannot reproduce the measured methanol
 abundances in IRDC G028.23-00.19 either, because the new cosmic
 ray-induced desorption rate for methanol is negligible 
\citep{Collings04,Garrod07}. Therefore, our observations of
 IRDC G028.23-00.19 show for the first time that the exothermicity of
 surface addition reactions can also explain the observed methanol
 abundances of a massive, IR-quiescent clump (see this process included in 
a chemical network applied to IRDCs in the recent paper of Vasyunina et al.,
 submitted). 

\subsection{NH$_2$D and C$_2$H: cold gas tracers}
\label{deuterated}

\subsubsection{NH$_2$D}

Deuterated molecules are highly enhanced in the cold gas of prestellar cores,
 with respect to the local interstellar D/H value of 2.3\x10$^{-5}$
 \citep{Linsky06}. Enhancements of the NH$_2$D/NH$_3$ abundance ratio as
 high as 0.7 \citep{Pillai07} and 0.8 \citep{Busquet10} have been measured
 in prestellar massive cores. In cold gas, the deuterium enrichment is
 primarily initiated by the reaction H$^+_3$\,+\,HD $\Longleftrightarrow$
 H$_2$D$^+$\,+\,H$_2$\,+\,$\Delta E$, which is exothermic by
 $\Delta E/k=230$ K \citep{Millar89}. At the typical temperatures of prestellar
 cores ($\lesssim$20 K), the reverse reaction becomes negligible and the 
H$_2$D$^+$/H$^+_3$ abundance ratio (and others molecular D/H ratios) is  
larger than the interstellar D/H. The degree of
 deuteration is even higher when CO (the main destroyer of H$^+_3$ and 
H$_2$D$^+$) is removed from the gas phase \citep[e.g.,][]{Chen11,Crapsi05}.

Among all the observed molecules in this work, NH$_2$D has the narrowest 
 profile with a line width of 1.3 \kms\ at the peak position. The NH$_2$D
 abundance of 1.3\x10$^{-9}$ is high and comparable to the usually more
 abundant HCO$^+$ and HNC, and suggests that the deuterated fraction is
 high. However, without NH$_3$ observations we cannot determine the exact 
value of the deuterated fraction. Because NH$_2$D emission is
 confined to the central part of the clump MM1 (see Figure~\ref{four_maps2}),
 this region is probably the coldest and densest part of the IRDC. 

\subsubsection{C$_2$H}

C$_2$H has been known to be a PDR tracer \cite[e.g.,][]{Fuente93}.
 However, \cite{Beuther08} suggest that this molecule could also be used 
to study the cold gas associated with the initial conditions of high-mass
 star formation. 
 \cite{Sanhueza12} detect C$_2$H in $\sim$30\% of their sub-sample composed 
by IR quiescent clumps. They suggest that 
mapping the C$_2$H distribution at high angular resolution may help to 
clarify if the C$_2$H emission comes from the external photodissociated 
layers of clumps or if the C$_2$H emission comes from the dense, cold gas
 inside IR quiescent clumps. 

As can be seen in Figure~\ref{four_maps2}, the spatial distribution of 
the C$_2$H emission resembles that of NH$_2$D. In IRDC G028.23-00.19,  
the C$_2$H emission comes from the central and coldest part of the cloud, 
and not from the external layers of the IRDC. C$_2$H and NH$_2$D molecules have 
their peak emission in the same place and their spatial extensions are similar 
(although C$_2$H extends slightly further out of the clump MM1). Because 
C$_2$H emission is mostly located at the same position as the NH$_2$D
 emission, i.e. the center of MM1, we suggest C$_2$H is also tracing dense,
 cold gas in IRDC G028.23-00.19. 

\subsection{HCO$^+$, H$^{13}$CO$^+$, HNC, HN$^{13}$C, and N$_2$H$^+$}

HCO$^+$ and HNC abundances are estimated using their isotopologues
 (H$^{13}$CO$^+$ and HN$^{13}$C). The HCO$^+$ abundance of 
 1.2\x10$^{-9}$ and HNC abundance of 1.5\x10$^{-9}$ are consistent, within the 
uncertainties, with the values determined by \cite{Sanhueza12} in IRDC
 G028.23-00.19 MM1, using 12 K in 
their calculation. The abundances determined in the MM1 clump are about one
 order of magnitude lower than the typical values found by \cite{Sanhueza12}
 in their sample of $\sim$30 IR, quiescent clumps (median values of 
2.4\x10$^{-8}$ for HCO$^+$ and 3.5\x10$^{-8}$ for HNC). They found that
 HCO$^+$ and HNC abundances increase as the clumps evolve, from IR,
 quiescent clumps to protostellar objects and on to \hii\ regions. The low
 abundances measured in our work for the clump MM1 in IRDC G028.23-00.19
 suggest that this clump is in a very early stage of evolution. 

The relative intensities of the three hyperfine transitions of N$_2$H$^+$
 $J=1 \rightarrow0$ are 1:3:5, with the brightest
 line in the center (\tNdosHudt). However, toward 
the peak position of IRDC G028.23-00.19 (see Figure~\ref{spec_MM1}) and 
on a large region of the IRDC, the brightest hyperfine line is the one
 with the lowest relative intensity ratio (\tNdosHucu). This peculiar
 intensity ratio suggests a very high optical depth for N$_2$H$^+$. Thus, 
the N$_2$H$^+$ abundance of 4.3\x10$^{-10}$ is a lower limit.

\subsection{The role of the subclouds}

Because line widths in IRDCs are narrower than those in molecular clouds
 with copious star formation, IRDCs are excellent laboratories to study
 large sub-structures inside molecular clouds, such as subclouds or filaments.
 \cite{Jimenez10} and \cite{Devine11} also find close velocity
 components of $\lesssim$2 \kms\ in two different IRDCs. \cite{Jimenez10} find 
three subclouds in IRDC G035.39-00.33 using IRAM single-dish observations of 
C$^{18}$O. They find a widespread narrow SiO emission in this IRDC that
 could have been produced by the interaction of the subclouds. In another
 IRDC, G019.30+00.07, \cite{Devine11} find three subclouds using VLA
 observations of NH$_3$ and CCS. They find that 
 NH$_3$ and CCS trace different part of the subclouds: NH$_3$ peaks in the 
densest regions and CCS peaks in the subcloud-subcloud and outflow-subcloud 
interfaces. They suggest that in these boundary regions the gas is
 chemically young due to collisions that release molecules from dust grains 
to the gas phase, leading to the presence of early time molecules like CCS.

We find that two subclouds form IRDC G028.23-00.19. Notably, the narrow 
SiO component is located in the region where both subclouds overlap. Assuming 
that this SiO emission is produced by the collision of the subclouds, this 
work adds new evidence in favor of the idea that 
the subclouds or filaments recently found in IRDCs are interacting and 
are part of the same cloud, and not just a superposition 
of clouds along the line of sight. Thus, it is not simply a coincidence   
that IRDCs frequently show no evident signs of active star formation, IRDCs
 may be in a very early stage in which the clouds are still being assembled. 

\subsection{The ``prestellar'' nature of IRDC G028.23-00.19}

Except for the bright unrelated OH/IR star superimposed against 
the northern part of the 
 cloud \citep{Bowers89}, IRDC G028.23-00.19 is IR dark from 3.6 to 70 $\mu$m.  
The absence of embedded IR sources suggest that the whole IRDC is in an 
early stage of evolution, without signs of active star formation. However, 
our CARMA observations show broad SiO and CH$_3$OH emission in the northern
 and southern regions of the IRDC that may indicate molecular outflow
 activity and, consequently, current star formation. In addition, one of
 the possible explanations for the narrow SiO component in the center-west
 part of the IRDC is that the SiO emission is produced for an unresolved
 population of low-mass stars.

 On the other hand, the central clump MM1 is quiescent. It is cold (12 K) and
 lacks IR sources, cm continuum and SiO emission are not detected, molecular
 line widths are narrow ($\lesssim$2 \kms), and the NH$_2$D abundance is high.
 The combination of all these   
 characteristics of MM1 indicates that this clump is still in a prestellar 
stage. Although regions in the same IRDC may show indirect signs of 
star formation, all the observations are consistent with 
the massive clump MM1 being a pristine starless clump. To determine whether 
MM1 will form stars in the future, we compare the dust mass ($M_{\rm dust}$) 
with the virial mass ($M_{\rm vir}$). In Section~\ref{dust_temp} we estimated 
a $M_{\rm dust}$ of 1520 \Msun, using the 1.2 mm dust continuum emission.
 To determine the virial mass, we follow the prescription of
 \cite{MacLaren88}. The virial mass of a clump with an uniform density
 profile, neglecting magnetic fields and external forces, is given by
 $M_{\rm vir}=210\,R\,\Delta V^2$ \Msun, where $R$ is the radius of the clump
 in pc and $\Delta V$ is the line width in \kms. Adopting the same size used to 
obtain $M_{\rm dust}$, a radius of 0.6 pc, and an average line width for MM1
 of 1.9 \kms\ (from Table~\ref{tbl-gaussian}), we obtain a virial mass
 of $\sim$450 \Msun. Therefore, the virial parameter,
 $\alpha=M_{\rm vir}/M_{\rm dust}$, is 0.3, indicating that the clump MM1
 is gravitationally bound and eventually, due to its large mass, will
 form stars. To further investigate if the clump MM1 will form high-mass 
stars, we use the results from \cite{Kauffmann10}. They find an empirical 
massive star formation threshold, based on clouds with and without high-mass 
star formation. They suggest that IRDCs with masses larger than the mass 
limit given by $m_{\rm lim}$ = 870 \Msun\ (r/pc)$^{1.33}$ are forming high-mass 
stars or will form in the future. Applying this relationship to MM1, its 
corresponding mass limit is 440 \Msun. Thus, ``the compactness''
 ($M_{\rm dust}$/$m_{\rm lim}$) of MM1 is 3.5 and it is highly likely that the 
clump will form high-mass stars in the future. 

Our observation that other regions in the IRDC, but not MM1, may be forming
 stars raises a number of questions. Why would star formation not start first 
 in MM1? If we observe at higher angular resolution and sensitivity, would
 we be able to resolve and associate cores or protostars with the SiO emission?
 How does the scenario observed in IRDC G028.23-00.19 fit into current
 models of high-mass star formation? Such questions may be possibly answered
 with observations at higher angular resolution and better sensitivity. 

 \section{Conclusions}

We have observed the IRDC G028.23-00.19 in several molecular species (NH$_2$D,
 H$^{13}$CO$^+$, SiO, HN$^{13}$C, C$_2$H, HCO$^+$, HNC, N$_2$H$^+$, and CH$_3$OH) 
and continuum emission at 3.3 mm using CARMA (11\arcsec\ angular resolution).
  This IRDC is dark at
 {\it Spitzer} 3.6, 4.5, 8.0, and 24 $\mu$m and {\it Herschel} 70 $\mu$m.
 In its center, the IRDC hosts one of the most massive IR,
 quiescent clumps known (MM1). We have examined the spectral line observations 
and draw the following conclusions:

1. Using the emission detected at 250, 350, and 500 $\mu$m with {\it Herschel},
 and the 1.2 mm emission from IRAM 30 m, we updated the mass (1520 \Msun) and 
the dust temperature (12 K) of the central clump in IRDC G028.23-00.19.

2. The low temperature, high NH$_2$D abundance, low HCO$^+$ and HNC abundances, 
non-detection of SiO, narrow line widths, and absence of embedded IR
 sources in MM1 indicate that the clump still remains in a prestellar phase.

3. Strong SiO components with broad line widths are detected in the
northern and southern regions of the IRDC. The large line widths
 (6-9 \kms) and high abundances (0.8-3.8\x10$^{-9}$) of SiO suggest that
 the mechanism releasing SiO from the dust grains into the gas phase could
 be molecular outflows from undetected intermediate-mass stars or clusters  
of low-mass stars deeply embedded in the IRDC. However, in the southern
 region where the SiO abundance peaks, there is no associated counterpart
 in the continuum. 

4. A weaker SiO component is detected in the center-west part of
 the IRDC. The narrow line widths ($\sim$2 \kms) and low SiO abundances
 (4.3\x10$^{-11}$) are consistent with either a ``subcloud-subcloud''
 interaction or an unresolved population of a few low-mass stars. Higher
 angular resolution observations are needed to clearly establish the
 origin of the narrow SiO emission.
 
5. We report unexpected widespread CH$_3$OH emission throughout the whole
 IRDC and the first detection of extended methanol emission in a massive
 prestellar clump. Because IR observations show no primary protostar
 embedded in the IRDC, we reject thermal evaporation of methanol from
 dust mantles as the production mechanism. CH$_3$OH emission with broad
 line widths support the idea of molecular outflows in the southern and
 northern regions of the IRDC. The CH$_3$OH emission at the position of MM1
is rather narrow. We suggest that the most likely mechanism able to release
 methanol from the dust grains to the gas phase in such a cold region is the
exothermicity of surface reactions \citep{Garrod06,Garrod07}. 

6. C$_2$H has been suggested to be a PDR tracer. However, recent evidence  
suggests it could also trace cold, dense gas associated with earlier stages
 of star formation. Because the spatial distribution of C$_2$H emission
 resembles that of NH$_2$D, we suggest C$_2$H traces cold, dense gas in
 this IRDC. 

7. HN$^{13}$C reveals that the IRDC is composed of two substructures
 (``subclouds'') separated by 1.4 \kms. Remarkably, the
 subclouds overlap in the center-west region of the IRDC, exactly coincident  
with the narrow SiO component. One explanation for the 
narrow SiO component is that both subclouds may be interacting and
 producing low velocity shocks that release small amounts of SiO to
 the gas phase. We speculate that IRDCs may be young molecular clouds that 
could still be in a stage where they are being assembled. 

8. In the densest part of the IRDC, the MM1 clump shows no signs of star
 formation. Notably, in other regions of the IRDC it appers that star
 formation activity may be occurring. Why star formation would begin in less
 dense regions and how the findings on IRDC G028.23-00.19 fit into 
 current models of high-mass star formation are open questions that will
 be addressed in future investigations. 

\acknowledgements

P.S. gratefully acknowledge to the instructors of the CARMA Summer School
 2011, specially for their enthusiasm and dedication to the professors
 John Carpenter, Nikolaus Volgenau, Melvyn Wright, Dick Plambeck, and
 Marc Pound. P.S. thanks Andr{\'e}s Guzm{\'a}n for his careful 
 reading of the paper. We also thank the anonymous referee for helpful 
comments that improved the paper. P.S. and J.M.J. acknowledge funding
 support from NSF Grant No. AST-0808001 and AST-1211844. I.J-S. acknowledges
 the financial 
 support from the People Programme (Marie Curie Actions) of the European
 Union's Seventh Framework Programme (FP7/2007-2013) under REA grant
 agreement number PIIF-GA-2011-301538. 
Support for CARMA construction was derived from the states
 of California, Illinois, and Maryland, the James S. McDonnell Foundation,
 the Gordon and Betty Moore Foundation, the Kenneth T. and Eileen L. Norris
 Foundation, the University of Chicago, the Associates of the California
 Institute of Technology, and the National Science Foundation. Ongoing CARMA
 development and operations are supported by the National Science Foundation
 under a cooperative agreement, and by the CARMA partner universities.


\begin{thebibliography}{}

\bibitem[Andr{\'e} et al.(2009)]{Andre09} Andr{\'e}, P., Basu, 
S., \& Inutsuka, S.\ 2009, Structure Formation in Astrophysics, 254 
\bibitem[Battersby et al.(2010)]{Battersby10} Battersby, C., Bally, 
J., Jackson, J.~M., et al.\ 2010, \apj, 721, 222 
\bibitem[Benjamin et al.(2003)]{Benjamin03} Benjamin, R.~A., 
Churchwell, E., Babler, B.~L., et al.\ 2003, \pasp, 115, 953 
\bibitem[Beuther et al.(2008)]{Beuther08} Beuther, H., Semenov, 
D., Henning, T., \& Linz, H.\ 2008, \apjl, 675, L33
\bibitem[Brown \& Bolina(2007)]{Brown07} Brown, W.~A., \& Bolina,
 A.~S.\ 2007, \mnras, 374, 1006 
\bibitem[Bowers \& Knapp(1989)]{Bowers89} Bowers, P.~F., \& Knapp, G.~R.\
 1989, \apj, 347, 325 
\bibitem[Busquet et al.(2010)]{Busquet10} Busquet, G., Palau, A.,
 Estalella, R., et al.\ 2010, \aap, 517, L6 
\bibitem[Carey et al.(2009)]{Carey09} Carey, S.~J., 
Noriega-Crespo, A., Mizuno, D.~R., et al.\ 2009, \pasp, 121, 76 
\bibitem[Carey et al.(2000)]{Carey00} Carey, S.~J., Feldman, 
P.~A., Redman, R.~O., et al.\ 2000, \apjl, 543, L157
\bibitem[Carey et al.(1998)]{Carey98} Carey, S.~J., Clark, 
F.~O., Egan, M.~P., et al.\ 1998, \apj, 508, 721 
\bibitem[Caselli et al.(1997)]{Caselli97} Caselli, P., Hartquist, T.~W.,
 \& Havnes, O.\ 1997, \aap, 322, 296 
\bibitem[Chambers et al.(2009)]{Chambers09} Chambers, E.~T., 
Jackson, J.~M., Rathborne, J.~M., \& Simon, R.\ 2009, \apjs, 181, 360 
\bibitem[Chen et al.(2011)]{Chen11} Chen, H.-R., Liu, S.-Y., 
Su, Y.-N., \& Wang, M.-Y.\ 2011, \apj, 743, 196 
\bibitem[Chira et al.(2013)]{Chira13} Chira, R.-A., Beuther, H., Linz, H., et al.\ 2013, \aap, 552, A40 
\bibitem[Collings et al.(2004)]{Collings04} Collings, M.~P., 
Anderson, M.~A., Chen, R., et al.\ 2004, \mnras, 354, 1133 
\bibitem[Crapsi et al.(2005)]{Crapsi05} Crapsi, A., Caselli, P., 
Walmsley, C.~M., et al.\ 2005, \apj, 619, 379 
\bibitem[Daniel et al.(2006)]{Daniel06} Daniel, F., Cernicharo, J.,
 \& Dubernet, M.-L.\ 2006, \apj, 648, 461 
\bibitem[Devine et al.(2011)]{Devine11} Devine, K.~E., Chandler, 
C.~J., Brogan, C., et al.\ 2011, \apj, 733, 44 
\bibitem[Dickens et al.(2000)]{Dickens00} Dickens, J.~E., Irvine, 
W.~M., Snell, R.~L., et al.\ 2000, \apj, 542, 870 
\bibitem[Egan et al.(1998)]{Egan98} Egan, M.~P., Shipman, 
R.~F., Price, S.~D., et al.\ 1998, \apjl, 494, L199 
\bibitem[Fontani et al.(2011)]{Fontani11} Fontani, F., Palau, A., Caselli, P.,
 et al.\ 2011, \aap, 529, L7 
\bibitem[Friberg et al.(1988)]{Friberg88} Friberg, P., Hjalmarson, A.,
 Madden, S.~C., \& Irvine, W.~M.\ 1988, \aap, 195, 281 
\bibitem[Fuchs et al.(2009)]{Fuchs09} Fuchs, G.~W., Cuppen, H.~M.,
 Ioppolo, S., et al.\ 2009, \aap, 505, 629 
\bibitem[Fuente et al.(1993)]{Fuente93} Fuente, A., Martin-Pintado, J.,
 Cernicharo, J., \& Bachiller, R.\ 1993, \aap, 276, 473
\bibitem[Garay \& Lizano(1999)]{Garay99} Garay, G., \& Lizano, S.\ 1999,
 \pasp, 111, 1049 
\bibitem[Garden et al.(1991)]{Garden91} Garden, R.~P., Hayashi, 
M., Hasegawa, T., Gatley, I., \& Kaifu, N.\ 1991, \apj, 374, 540
\bibitem[Garrod et al.(2007)]{Garrod07} Garrod, R.~T.,
 Wakelam, V., \& Herbst, E.\ 2007, \aap, 467, 1103 
\bibitem[Garrod et al.(2006)]{Garrod06} Garrod, R., Park, I.~H., 
Caselli, P., \& Herbst, E.\ 2006, Faraday Discussions, 133, 51 
\bibitem[Girart et al.(2002)]{Girart02} Girart, J.~M., Viti, S., Williams,
 D.~A., Estalella, R., \& Ho, P.~T.~P.\ 2002, \aap, 388, 1004 
\bibitem[G{\'o}mez et al.(2011)]{Gomez11} G{\'o}mez, L., Wyrowski, F.,
 Pillai, T., Leurini, S., \& Menten, K.~M.\ 2011, \aap, 529, A161 
\bibitem[Green et al.(2009)]{Green09} Green, S.~D., Bolina, 
A.~S., Chen, R., et al.\ 2009, \mnras, 398, 357 
\bibitem[Hasegawa \& Herbst(1993)]{Hasegawa93} Hasegawa, T.~I., \& Herbst,
 E.\ 1993, \mnras, 261, 83 
\bibitem[Henshaw et al.(2013)]{Henshaw13} Henshaw, J.~D., 
Caselli, P., Fontani, F., et al.\ 2013, \mnras, 428, 3425 
\bibitem[Hoq et al.(2013)]{Hoq13} Hoq, S., Jackson, J.M., Foster, J.B., 
Sanhueza, P., Whitaker, J.S., Claysmith, C., Rathborne, J., Vasyunina, T., \& 
Vasyunin, A.I.\ 2013, ApJ, submitted
\bibitem[Jackson \& Kraemer(1999)]{Jackson99} Jackson, J.~M., \& Kraemer, K.~E.\ 1999, \apj, 512, 260 
\bibitem[Jim{\'e}nez-Serra et al.(2010)]{Jimenez10} Jim{\'e}nez-Serra, I.,
 Caselli, P., Tan, J.~C., et al.\ 2010, \mnras, 406, 187 
\bibitem[Jim{\'e}nez-Serra et al.(2008)]{Jimenez08} Jim{\'e}nez-Serra, I.,
 Caselli, P., Mart{\'{\i}}n-Pintado, J., \& Hartquist, T.~W.\ 2008, \aap,
 482, 549 
\bibitem[Jim{\'e}nez-Serra et al.(2005)]{Jimenez05} Jim{\'e}nez-Serra, I.,
 Mart{\'{\i}}n-Pintado, J., Rodr{\'{\i}}guez-Franco, A., \& Mart{\'{\i}}n,
 S.\ 2005, \apjl, 627, L121 
\bibitem[Kauffmann \& Pillai(2010)]{Kauffmann10} Kauffmann, J., \& Pillai,
 T.\ 2010, \apjl, 723, L7 
\bibitem[Kim et al.(2010)]{Kim10} Kim, G., Lee, C.~W., Kim, 
J., et al.\ 2010, Journal of Korean Astronomical Society, 43, 9 
\bibitem[Lefloch et al.(1998a)]{Lefloch98a} Lefloch, B., Castets, A., Cernicharo, J., Langer, W.~D., \& Zylka, R.\ 1998, \aap, 334, 269 
\bibitem[Lefloch et al.(1998b)]{Lefloch98b} Lefloch, B., Castets, 
A., Cernicharo, J., \& Loinard, L.\ 1998, \apjl, 504, L109 
\bibitem[Linsky et al.(2006)]{Linsky06} Linsky, J.~L., Draine, 
B.~T., Moos, H.~W., et al.\ 2006, \apj, 647, 1106 
\bibitem[MacLaren et al.(1988)]{MacLaren88} MacLaren, I., 
Richardson, K.~M., \& Wolfendale, A.~W.\ 1988, \apj, 333, 821 
\bibitem[Markwardt(2009)]{Markwardt09} Markwardt, C.~B.\ 2009, 
Astronomical Data Analysis Software and Systems XVIII, 411, 251
\bibitem[Martin-Pintado et al.(1992)]{Martin92} Martin-Pintado,
 J., Bachiller, R., \& Fuente, A.\ 1992, \aap, 254, 315 
\bibitem[Miettinen et al.(2011)]{Miettinen11} Miettinen, O., Hennemann, M.,
 \& Linz, H.\ 2011, \aap, 534, A134 
\bibitem[Milam et al.(2005)]{Milan05} Milam, S.~N., Savage, C., 
Brewster, M.~A., Ziurys, L.~M., \& Wyckoff, S.\ 2005, \apj, 634, 1126 
\bibitem[Millar et al.(1989)]{Millar89} Millar, T.~J., Bennett, 
A., \& Herbst, E.\ 1989, \apj, 340, 906 
\bibitem[Molinari et al.(2010)]{Molinari10} Molinari, S., Swinyard, B.,
 Bally, J., et al.\ 2010, \aap, 518, L100 
\bibitem[M{\"u}ller et al.(2005)]{Muller05} M{\"u}ller, H.~S.~P.,
 Schl{\"o}der, F., Stutzki, J., \& Winnewisser, G.\ 2005, Journal of
 Molecular Structure, 742, 215 
\bibitem[M{\"u}ller et al.(2001)]{Muller01} M{\"u}ller, H.~S.~P., Thorwirth,
 S., Roth, D.~A., \& Winnewisser, G.\ 2001, \aap, 370, L49 
\bibitem[{\"O}berg et al.(2005)]{Oberg05} {\"O}berg, K.~I., van 
Broekhuizen, F., Fraser, H.~J., et al.\ 2005, \apjl, 621, L33 
\bibitem[Ossenkopf \& Henning(1994)]{Ossenkopf94} Ossenkopf, V., \& Henning,
 T.\ 1994, \aap, 291, 943 
\bibitem[Perault et al.(1996)]{Perault96} Perault, M., Omont, A., Simon, G., et 
al.\ 1996, \aap, 315, L165
\bibitem[Peretto \& Fuller(2009)]{Peretto09} Peretto, N., \& Fuller,
 G.~A.\ 2009, \aap, 505, 405
\bibitem[Pickett et al.(1998)]{Pickett98} Pickett, H.~M., Poynter, R.~L.,
 Cohen, E.~A., et al.\ 1998, \jqsrt, 60, 883 
\bibitem[Pillai et al.(2006)]{Pillai06} Pillai, T., Wyrowski,
 F., Carey, S.~J., \& Menten, K.~M.\ 2006, \aap, 450, 569 
\bibitem[Pillai et al.(2011)]{Pillai11} Pillai, T., Kauffmann, J.,
 Wyrowski, F., et al.\ 2011, \aap, 530, A118
\bibitem[Pillai et al.(2007)]{Pillai07} Pillai, T., Wyrowski, F., Hatchell,
 J., Gibb, A.~G., \& Thompson, M.~A.\ 2007, \aap, 467, 207 
\bibitem[Pillai et al.(2006)]{Pillai06} Pillai, T., Wyrowski,
 F., Carey, S.~J., \& Menten, K.~M.\ 2006, \aap, 450, 569 
\bibitem[Pineda \& Teixeira(2013)]{Pineda13} Pineda, J.~E, \&
 Teixeira, P.~S.\ 2013, arXiv:1305.3329 
\bibitem[Ragan et al.(2011)]{Ragan11} Ragan, S.~E., Bergin, 
E.~A., \& Wilner, D.\ 2011, \apj, 736, 163
\bibitem[Rathborne et al.(2006)]{Rathborne06} Rathborne, J.~M., 
Jackson, J.~M., \& Simon, R.\ 2006, \apj, 641, 389 
\bibitem[Rathborne et al.(2008)]{Rathborne08} Rathborne, J.~M., 
Jackson, J.~M., Zhang, Q., \& Simon, R.\ 2008, \apj, 689, 1141
\bibitem[Rathborne et al.(2010)]{Rathborne10} Rathborne, J.~M., 
Jackson, J.~M., Chambers, E.~T., Stojimirovic, I., Simon, R., Shipman, R., 
\& Frieswijk, W.\ 2010, \apj, 715, 310 
\bibitem[Requena-Torres et al.(2007)]{Requena07} Requena-Torres, 
M.~A., Marcelino, N., Jim{\'e}nez-Serra, I., et al.\ 2007, \apjl, 655, L37 
\bibitem[Sakai et al.(2012)]{Sakai12} Sakai, T., Sakai, N., 
Furuya, K., et al.\ 2012, \apj, 747, 140 
\bibitem[Sakai et al.(2010)]{Sakai10} Sakai, T., Sakai, N., 
Hirota, T., \& Yamamoto, S.\ 2010, \apj, 714, 1658 
\bibitem[Sakai et al.(2008)]{Sakai08} Sakai, T., Sakai, N., 
Kamegai, K., Hirota, T., Yamaguchi, N., Shiba, S., 
\& Yamamoto, S.\ 2008, \apj, 678, 1049 
\bibitem[Sanhueza et al.(2012)]{Sanhueza12} Sanhueza, P., Jackson, 
J.~M., Foster, J.~B., et al.\ 2012, \apj, 756, 60 
\bibitem[Sanhueza et al.(2010)]{Sanhueza10} Sanhueza, P., Garay, 
G., Bronfman, L., et al.\ 2010, \apj, 715, 18
\bibitem[Schilke et al.(1997)]{Schilke97} Schilke, P., Walmsley, C.~M.,
 Pineau des Forets, G., \& Flower, D.~R.\ 1997, \aap, 321, 293 
\bibitem[Simon et al.(2006a)]{Simon06a} Simon, R., Jackson, 
J.~M., Rathborne, J.~M., \& Chambers, E.~T.\ 2006, \apj, 639, 227
\bibitem[Su et al.(2007)]{Su07} Su, Y.-N., Liu, S.-Y., Chen, 
H.-R., Zhang, Q., \& Cesaroni, R.\ 2007, \apj, 671, 571 
\bibitem[Tielens et al.(1994)]{Tielens94} Tielens, A.~G.~G.~M., 
McKee, C.~F., Seab, C.~G., \& Hollenbach, D.~J.\ 1994, \apj, 431, 321 
\bibitem[van der Tak et al.(2000)]{Van00} van der Tak, F.~F.~S., van Dishoeck, E.~F., \& Caselli, P.\ 2000, \aap, 361, 327 
\bibitem[Vasyunina et al.(2011)]{Vasyunina11} Vasyunina, T., Linz,
 H., Henning, T., Zinchenko, I., Beuther, H., \& Voronkov, M.\ 2011,
 \aap, 527, A88 
\bibitem[Wang et al.(2006)]{Wang06} Wang, Y., Zhang, Q., 
Rathborne, J.~M., Jackson, J., \& Wu, Y.\ 2006, \apjl, 651, L125
\bibitem[Wang et al.(2011)]{Wang11} Wang, K., Zhang, Q., Wu, 
Y., \& Zhang, H.\ 2011, \apj, 735, 64 
\bibitem[Watanabe \& Kouchi(2002)]{Watanabe02} Watanabe, N., \& Kouchi,
 A.\ 2002, \apjl, 571, L173 
\bibitem[Wirstr{\"o}m et al.(2011)]{Wirstrom11} Wirstr{\"o}m, E.~S.,
 Geppert, W.~D., Hjalmarson, {\AA}., et al.\ 2011, \aap, 533, A24 
\bibitem[Zhang et al.(2009)]{Zhang09} Zhang, Q., Wang, Y., Pillai, T.,
 \& Rathborne, J.\ 2009, \apj, 696, 268


\end{thebibliography}
\end{document}